\documentclass{aa}
\usepackage{graphicx}
\usepackage{amssymb}
\usepackage{amsmath}
\usepackage{natbib}
\usepackage{mathrsfs}
\usepackage{ulem}
\usepackage{txfonts}
\usepackage{lmodern}

\newcommand{\corot}{{\textsc{CoRoT}}}

\newcommand{\kepler}{\textit{Kepler}}

\newcommand{\omc}{\langle\Omega\rangle_{\rm c}}
\newcommand{\ome}{\langle\Omega\rangle_{\rm e}}
\newcommand{\omtot}{\langle\Omega\rangle}

\newcommand{\ind}[1]{_{\rm #1}}

\def\m2s2{\,m$^{2}$\,s$^{-2}$} 

\newcommand{\vect}[1]{\boldsymbol{\rm #1}}

\newcommand{\vaisala}{Brunt-V\"ais\"al\"a}
\newcommand{\cesam}{\textsc{cesam2k}}

\newenvironment{itemize*}%
  {\begin{itemize}%
    \setlength{\itemsep}{1pt}%
    \setlength{\parskip}{1pt}}%
  {\end{itemize}}

\newcommand\avoir[1]{{#1}}

\newcommand\T{\rule{0pt}{2.6ex}}
\newcommand\B{\rule[-1.2ex]{0pt}{0pt}}

\usepackage[switch]{lineno} 

\begin{document}
\title{Near-degeneracy effects on the frequencies of rotationally-split mixed modes in red giants}
\titlerunning{Near-degeneracy effects on the frequencies of rotationally-split mixed modes in red giants}
\author{
S. Deheuvels\inst{1,2}
\and R.~M. Ouazzani\inst{3,4,5}
\and S. Basu \inst{6}
}


\institute{Universit\'e de Toulouse; UPS-OMP; IRAP; Toulouse, France
	\and CNRS; IRAP; 14, avenue Edouard Belin, F-31400 Toulouse, France	
	\and Institut d'Astrophysique et de G\'eophysique de l'Universit\'e de Li\`ege, All\'ee du 6 Ao\^ut 17, 4000 Li\`ege, Belgium
	\and Stellar Astrophysics Centre, Department of Physics and Astronomy, Aarhus University, Ny Munkegade 120, DK-8000 Aarhus C, Denmark
	\and Observatoire de Paris, LESIA, CNRS UMR 8109, 92195 Meudon, France
	\and Department of Astronomy, Yale University, P.O. Box 208101, New Haven, CT 06520-8101, USA
}

\offprints{S. Deheuvels\\ \email{sebastien.deheuvels@irap.omp.eu}
}

\date{Submitted ...}

\abstract{The \kepler\ space mission has made it possible to measure the rotational splittings of mixed modes in red giants, thereby providing an unprecedented opportunity to probe the internal rotation of these stars.}
{Asymmetries have been detected in the rotational multiplets of several red giants. This is unexpected since all the red giants whose rotation have been measured thus far are found to rotate slowly, and low rotation, in principle, produces  symmetrical multiplets. Our aim here is to explain these asymmetries and find a way of exploiting them to probe the internal rotation of red giants.}
{We show that in the cases where asymmetrical multiplets were detected, near-degeneracy effects are expected to occur, because of the combined effects of rotation and mode mixing. Such effects have not been taken into account so far. By using both perturbative and non-perturbative approaches, we show that near-degeneracy effects produce multiplet asymmetries that are very similar to the observations. We then propose and validate a method based on the perturbative approach to probe the internal rotation of red giants using multiplet asymmetries.}
{We successfully apply our method to the asymmetrical $l=2$ multiplets of the \kepler\ young red giant KIC7341231 and obtain precise estimates of its mean rotation in the core and the envelope. The observed asymmetries are reproduced with a good statistical agreement, which confirms that near-degeneracy effects are very likely the cause of the detected multiplet asymmetries.}
{We expect near-degeneracy effects to be important for $l=2$ mixed modes all along the red giant branch (RGB). For $l=1$ modes, these effects can be neglected only at the base of the RGB. They must therefore be taken into account when interpreting rotational splittings and as shown here, they can bring valuable information about the internal rotation of red giants.}


\keywords{Stars: oscillations -- Stars: rotation -- Stars: evolution -- Stars: individual (KIC7341231)}

%

\maketitle

\section{Introduction \label{sect_intro}}

Rotation is known to have a large influence on the structure and the evolution of stars. In particular, it is expected to induce additional mixing of the chemical elements inside stars. Taking the effects of rotation into account is believed to be a key step to making progress in stellar modeling. However, the internal rotation profiles of stars and the way they are modified during their evolution remains uncertain. Rotation-induced processes of angular moment transport such as meridional circulation and shear instabilities (as they are currently modeled) are not efficient enough to account for the nearly solid-body rotation in the solar radiative interior that was found by helioseismologic analyses (\citealt{schou98}, \citealt{chaplin99}, \citealt{effdarwich13}). Other types of processes are likely to operate. The main candidates invoked so far are internal gravity waves (\citealt{charbonnel05}, \citealt{talon08},\citealt{fuller14}) and magnetic fields that either have a fossil origin (e.g. \citealt{gough98}, \citealt{spada10}, \citealt{maeder14}) or result from MHD instabilities (e.g. \citealt{spruit99}, \citealt{rudiger15}). However, their relative importance and the timescales over which they operate is still a matter of active debate.


The seismology of subgiants and red giants has recently proved that it could significantly contribute to understanding angular momentum transport inside stars. 
Owing to the large core density of red giants, the frequencies of gravity modes in these stars are comparable to the frequencies of the acoustic modes stochastically excited in the convective envelope. This gives rise to mixed modes, which behave as gravity modes in the core and as pressure modes in the envelope. Such modes were initially found in stellar models by \cite{dziembowski71}, and later \cite{scuflaire74}, \cite{osaki75}. They were first detected from the ground (\citealt{kjeldsen95b}). More recently,  with the advent of space missions \corot\ (\citealt{baglin06}) and \kepler\ (\citealt{borucki10}), mixed modes were detected in thousands of subgiants (\citealt{analyse_49385} \citealt{campante11}) and red giants (\citealt{bedding11}, \citealt{mosser11}). The measurement of rotational splittings of mixed modes (\citealt{beck12}, \citealt{deheuvels12}, \citealt{mosser12b}) has given direct insight into the internal rotation of red giants and their variations with evolution. In the subgiant phase, \cite{deheuvels14} showed that the core spins up and the envelope spins down. This was qualitatively expected, considering that the deepest layers below the H-burning shell contract, while the layers above expand. However, the core rotation rates measured with seismology are several orders of magnitude below the values predicted by theoretical models that include rotation-induced transport of angular momentum (\citealt{eggenberger12}, \citealt{marques13}). On the red giant branch (RGB), seismic measurements of the core rotation for several hundreds of \kepler\ targets led to the striking observation that the core of red giants is in fact spinning down (\citealt{mosser12b}), which is also at odds with current theoretical models. These results are clear evidence that an efficient redistribution of angular momentum between the core and the envelope occurs in red giants, the origin of which remains unknown. More recently, it has been shown that a similar conclusion can be drawn for intermediate-mass stars in the core-He burning phase, where an even milder radial differential rotation was observed, with core-envelope ratios around two (\citealt{deheuvels15}). These novel observational constraints prompted new theoretical studies about the efficiency of angular momentum transport caused by internal gravity waves (\citealt{fuller14}), MHD instabilities (\citealt{rudiger15}, \citealt{jouve15}), or mixed modes themselves \citep{belkacem15a,belkacem15b}, but the question remains open.

So far, all the red giants whose rotation profile could be seismically measured have been found to be slow rotators; they rotate in a regime where the effects of the centrifugal force on stellar pulsations can be safely neglected and the rotation frequency remains small compared the pulsation frequencies. In such case, the effects of rotation can be treated as a first-order perturbation to the equations of non-radial oscillations. For a spherically-symmetric rotation profile $\Omega(r)$, the frequency shift for a mode of radial order $n$, degree $l$, and azimuthal order $m$ can be expressed as 
\begin{linenomath*}
\begin{equation}
\delta\omega_{n,l,m} \approx m \int_0^R K_{n,l}(r) \Omega(r) \,\hbox{d}r.
\label{eq_split_inv}
\end{equation}
\end{linenomath*}
where the functions $K_{n,l}(r)$ are the rotational kernels. According to Eq. \ref{eq_split_inv}, the components of a rotational multiplet are expected to be uniformly spaced, and thus symmetrical with respect to the central $m=0$ component. While most rotational multiplets in red giants indeed show this symmetry, asymmetries have been reported in several \kepler\ red giants. \cite{deheuvels12} found significant asymmetries in the rotational multiplets of two neighboring $l=2$ modes of the young red giant KIC7341231. \avoir{Asymmetries were also reported in the multiplets of several $l=1$ modes in the spectra of the red giants KIC5006817 (\citealt{beck14}) and KIC4448777 (\citealt{dimauro16}).}
These asymmetries need to be understood because they are at odds with what is expected from Eq. \ref{eq_split_inv}, which has been assumed by all the studies that have derived information about the internal rotation of red giants. 

In this paper, we argue that for KIC7341231 (and potentially also for the two other aforementioned targets), these asymmetries are caused by near-degeneracy effects, which occur because of the combined effects of rotation and mode mixing. So far, it has always been assumed that the effects of rotation on the frequencies of mixed modes could be estimated in the same manner as for regular modes. However, mode mixing occurs in red giants when p modes and g modes 
have frequencies close enough for them to couple through the evanescent region separating the two cavities. It is well known that when two modes have close frequencies, near-degeneracy effects may arise, which modify the frequency corrections to be applied (e.g., \citealt{dziembowski92}, \citealt{suarez06}). If the frequency spacing between mixed modes is of the order or smaller than the rotation rate, we expect near-degeneracy effects to become important. This issue has not been addressed so far for red giants. In this paper, we study this issue using KIC7341231 as a test case. We show that near-degeneracy effects need to be taken into account in this star and we demonstrate that they produce multiplet asymmetries that can account for the observations. 

The paper is organized as follows. In Sect. \ref{sect_obs}, we confirm and update the measurement of multiplet asymmetries in the oscillation spectrum of KIC7341231 using the full \kepler\ dataset and we demonstrate /that near-degeneracy effects cannot be neglected. In Sect. \ref{sect_pert}, we include the effects of near-degeneracy in the first-order perturbative treatment of rotation in the oscillation equations. We show that this correction indeed produces asymmetrical multiplets that are in qualitative agreement with the observations, and we argue that the intensity of the asymmetries depends on the level of radial differential rotation and on the trapping of the modes. In Sect. \ref{sect_acor}, we use the non-perturbative oscillation code \textsc{ACOR} (\citealt{ouazzani12}) to validate the first-order perturbative approach. In Sect. \ref{sect_diffrot}, we propose a new method to measure the average core and envelope rotation rates using the asymmetries in $l=2$ rotational multiplets. This method is thoroughly tested and validated using simulated data, and then successfully applied to KIC7341231 in Sect. \ref{sect_otto}. Sect. \ref{sect_discussion} is dedicated to a discussion of the results and we conclude in Sect. \ref{sect_conclusion}.

\section{Asymmetry of rotational multiplets in \kepler\ data: the test-case of KIC7341231 \label{sect_obs}}


\begin{figure}
\begin{center}
\includegraphics[width=9cm]{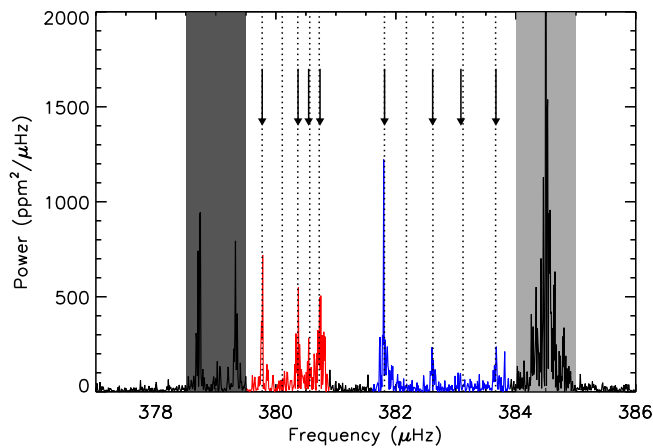}
\end{center}
\caption{Section of the oscillation spectrum of KIC7341231 in the neighborhood of the 384-$\mu$Hz radial mode (light grey area). The dark grey area corresponds to a rotationally-split $l=1$ mode and the vertical arrows indicate the observed mode frequencies of two $l=2$ mixed modes (see Table \ref{tab_otto_freq}). The vertical dotted lines show the results of a fit of Eq. \ref{eq_pert_omce} to the observed mode frequencies.
\label{fig_avcross_l2_otto}}
\end{figure}

\begin{table}
  \begin{center}
  \caption{Extracted parameters for the components of two $l=2$ multiplets in avoided crossing in the spectrum of KIC7341231. \label{tab_otto_freq}}
\begin{tabular}{c c c c} 
\hline \hline
\T \B $m$ & $\nu$ ($\mu$Hz) & $H$ (ppm$^2/\mu$Hz) & $\Gamma$ ($\mu$Hz) \\
\hline
\T -2 & $379.773 \pm 0.008$ & $ 340.0 \pm  242.7$ & $ 0.04 \pm  0.02$ \\
 0 & $380.372 \pm 0.013$ & $ 220.7 \pm  123.4$ & $ 0.07 \pm  0.03$ \\
 1 & $380.547 \pm 0.012$ & $ 124.2 \pm  174.7$ & $ 0.03 \pm  0.05$ \\
 2 & $380.741 \pm 0.015$ & $ 288.4 \pm  125.7$ & $ 0.11 \pm  0.04$ \\
-2 & $381.813 \pm 0.011$ & $ 385.6 \pm  186.7$ & $ 0.08 \pm  0.02$ \\
 0 & $382.615 \pm 0.010$ & $ 142.1 \pm  101.2$ & $ 0.05 \pm  0.02$ \\
1 & $383.085 \pm 0.052$ & $  27.2 \pm   20.0$ & $ 0.17 \pm  0.18$ \\
\B 2 & $383.669 \pm 0.014$ & $ 112.1 \pm   87.0$ & $ 0.07 \pm  0.06$ \\
\hline
\end{tabular}
\end{center}
\end{table}

The star KIC7341231 is a young red giant, which was observed with the \kepler\ spacecraft in short cadence (58.84876 s) from quarter Q5 through quarter Q17 of the mission. This represents 3.14 yr of nearly continuous data (duty cycle of 91\%). Based on the data from quarters Q5 through Q8 (about one year of data), \cite{deheuvels12} could measure the rotational splittings of 17 $l=1$ modes and 2 $l=2$ modes. By using the OLA (Optimally Localized Averages) inversion method, they obtained a precise estimate of the mean core rotation ($\omc/2\pi=710\pm51$ nHz). The authors showed that the measured splittings were however do not allow to build an average kernel for the envelope that would efficiently suppress the core contribution, and they were thus only able to derive an upper limit for the envelope mean rotation rate ($\ome/2\pi<150\pm19$ nHz). This limitation is inherent to the eigenfunctions of the detected modes and the increased precision to which the rotational splittings could be measured with the complete \kepler\ dataset would not solve this problem. It might however be that the rotational splittings of additional modes have become measurable, which might help canceling the core contribution in the envelope average kernel. This is out of the scope of the present study. 

\cite{deheuvels12} reported clear asymmetries for two closely-spaced $l=2$ mixed modes in the spectrum of KIC7341231 using the one-year-long observation available at that time. For this reason, these two modes were at that time excluded from the dataset used for rotation inversions. This is unsatisfactory since these modes clearly carry information on the internal rotation of the star. Additionally, since the coupling between the p-mode and g-mode cavities is weaker for $l=2$ modes, these modes could provide more precise estimate of the envelope rotation. This study is aimed at understanding these asymmetries in rotational multiplets and we used KIC7341231 as a test-case.

We updated the measurements of multiplet asymmetries in this star using the full-length \kepler\ data (quarters Q5 through Q17). Fig. \ref{fig_avcross_l2_otto} shows a section of the power spectrum of the star in the neighborhood of the two $l=2$ modes. The $m=(-2,0,2)$ components are the most prominent because the inclination angle of the star is close to $90^\circ$; only these components had been detected by \cite{deheuvels12} with the observation available at that time. When using the full \kepler\ data, we find that the $m=+1$ components of both $l=2$ multiplets also stand out significantly above the noise. We extracted the mode parameters (frequency, height, width) of all the components of the multiplets using the fitting technique described in \cite{deheuvels12}. The results are summarized in Table \ref{tab_otto_freq}. The detected components of the two multiplets, indicated by arrows in Fig. \ref{fig_avcross_l2_otto}, show clear asymmetries.

The asymmetry of the multiplets can be conveniently quantified by computing the quantity
\begin{linenomath*}
\begin{equation}
\delta\ind{asym} \equiv \frac{\omega_{-m}+\omega_{+m}-2\omega_{0}}{\omega_{+m}-\omega_{-m}}
\label{eq_asym}
\end{equation}
\end{linenomath*}
With this definition, $\delta\ind{asym}=0$ for a symmetric multiplet, while $|\delta\ind{asym}|=1$ if either $\omega_{-m}$ or $\omega_{m}$ overlaps with the $\omega_0$ mode. A positive value for $\delta\ind{asym}$ means that the splittings of the $m>0$ components are larger than those of the $m<0$ components (and conversely for $\delta\ind{asym}<0$). For the two $l=2$ mixed modes of KIC7341231, we found $\delta\ind{asym}\sim-0.24\pm0.03$ (red mode in Fig. \ref{fig_avcross_l2_otto}) and $\delta\ind{asym}\sim0.14\pm0.01$ (blue mode in Fig. \ref{fig_avcross_l2_otto}) using the $m=\pm2$ components. 

The two multiplets that show asymmetries in the spectrum of KIC7341231 have close frequencies, which means that near-degeneracy effects may arise, as mentioned in Sect. \ref{sect_intro}. Using Table \ref{tab_otto_freq}, we find that the frequency spacing between the $m=0$ components of the two multiplets is $2.24\pm0.02\,\mu$Hz. The core rotation rate of KIC7341231 as obtained by \cite{deheuvels12} thus amounts to $\sim32$\% of the frequency spacing between the two modes, which is not negligible. Consequently, we expect near-degeneracy effects to arise and they need to be taken into account. In the following, we show that near-degeneracy effects can indeed produce multiplet asymmetries very similar to the observed ones.

For the sake of completeness, in Sect. \ref{sect_discussion} we mention other possible causes for multiplet asymmetries, which we find less likely for KIC734123.

%

\section{First-order perturbative approach including near-degeneracy \label{sect_pert}}

We tested the influence of near-degeneracy effects on rotational multiplets by including them in a first-order perturbative treatment of rotation. This has already been studied in detail by previous authors (e.g., \citealt{dziembowski92}, \citealt{suarez06}, \citealt{ouazzani12b}, see also \citealt{goupil09} for a review) so we briefly recall the procedure and refer to these authors for more details. 

\subsection{Non-degenerate frequencies}

The oscillation equations including the first order effects of rotation can be written as
\begin{linenomath*}
\begin{equation}
(\mathcal{L}_0+\mathcal{L}_1) \vect{\xi} - \omega^2  \vect{\xi} = 0
\label{eq_pert}
\end{equation}
\end{linenomath*}
where $\omega$ corresponds to the mode eigenfrequency, and $ \vect{\xi}$ is the associated eigenfunction. The first-order correction $\mathcal{L}_1$ to the non-rotating operator $\mathcal{L}_0$ includes the effect of the advection relative to the inertial frame and the contribution from the Coriolis force. The two operators are given by
\begin{linenomath*}
\begin{align}
\mathcal{L}_0\vect{\xi} & = \frac{\nabla p'}{\rho_0} - \vect{g}' - \frac{\rho'}{\rho_0}\vect{g}_0 \\
\mathcal{L}_1\vect{\xi} & = 2\omega (m\Omega - i\vect{\Omega}\times) \vect{\xi}
\end{align}
\end{linenomath*}
For simplicity, in the following the mode eigenfunctions are normalized by the mode inertias, so that $\langle\xi_{a}|\xi_{a}\rangle = 1$, where the inner product is defined as
\begin{linenomath*}
\begin{equation}
\langle\xi_{a}|\xi_{b}\rangle \equiv \int_0^R \rho_0 r^2 \left[ \xi_{{\rm r}, a}\xi_{{\rm r}, b} + L^2\xi_{{\rm h}, a}\xi_{{\rm h},b} \right] \hbox{d}r 
\end{equation}
\end{linenomath*}
and $L^2=l(l+1)$.

In the general case without near-degeneracy, the first-order correction $\omega_1$ to the mode eigenfrequency is given by the variational principle, and therefore
\begin{linenomath*}
\begin{equation}
\omega_{1} = \frac{1}{2\omega_{0}} \langle \mathcal{L}_1 \vect{\xi}_{0} | \vect{\xi}_{0} \rangle
\label{eq_om1a}
\end{equation}
\end{linenomath*}
If we further assume a spherically symmetric rotation profile $\Omega(r)$, Eq. \ref{eq_om1a} reduces to 
\begin{linenomath*}
\begin{equation}
\omega_{1} = m \int_0^R K(r)  \Omega(r) \,\hbox{d}r
\label{eq_om1}
\end{equation}
\end{linenomath*}
where the rotational kernel $K(r)$ is given by
\begin{linenomath*}
\begin{equation}
K(r) = \rho_0 r^2 \left[ \xi_{{\rm r},0}^2 + (L^2-1)\xi_{{\rm h},0}^2 - 2\xi_{{\rm r},0}\xi_{{\rm h},0} \right] 
\label{eq_K1}
\end{equation}
\end{linenomath*}
Eq. \ref{eq_om1} shows that without near-degeneracy, the splittings vary linearly with $m$, so that the components of a rotational multiplet are expected to be uniformly spaced, and thus symmetrical with respect to the central $m=0$ component. 

\subsection{Near-degenerate frequencies \label{sect_neardeg}}

We now consider the case of two modes with same degree $l$, same azimuthal order $m$, and near-degenerate frequencies,
i.e., $|\omega_{0, a}-\omega_{0, b}|\sim\Omega$, where the subscripts $a$ and $b$ refer to the two modes. In this case, the the modes are coupled and corrections need to be included in the perturbative calculation of mode frequencies. To first-order, the eigenfunctions of near-degenerate modes can be written as 
\begin{linenomath*}
\begin{equation}
\vect{\xi}  = A\vect{\xi}_{0,a} + B\vect{\xi}_{0,b}.
\label{eq_eigenfunction}
\end{equation}
\end{linenomath*}
 As described in Appendix \ref{app_neardeg}, the perturbed eigenfrequencies are then given by the following expression
\begin{linenomath*}
\begin{equation}
\omega = \frac{\omega_{a}+\omega_{b}}{2} \pm \frac{1}{2} \sqrt{(\omega_{a}-\omega_{b})^2 + 4\omega_{1,ab}^2} \label{eq_freq_pert}
\end{equation}
\end{linenomath*}
where $\omega_{a}$ and $\omega_{b}$ correspond to the first-order perturbed frequencies of modes $a$ and $b$ when near-degeneracy effects are ignored, e.g., for mode $a$
\begin{linenomath*}
\begin{equation}
\omega_a \equiv \omega_{0,a} + \omega_{1,a}
\label{eq_omab}
\end{equation}
\end{linenomath*}
The quantity $\omega_{1,ab}$ represents the coupling between the two modes. 
For a spherically symmetric rotation profile, we have
\begin{linenomath*}
\begin{equation}
\omega_{1,ab} =  m \int_0^R K_{ab}(r)  \Omega(r) \,\hbox{d}r,
\label{eq_om1ab}
\end{equation}
\end{linenomath*}
where
\begin{linenomath*}
\begin{align}
K_{ab}(r) =  \rho_0 r^2 & \left[    \xi_{{\rm r},0,a}\xi_{{\rm r},0,b} + (L^2-1)\xi_{{\rm h},0,a}\xi_{{\rm h},0,b} \right. \nonumber \\
& \left.  - \xi_{{\rm r},0,a}\xi_{{\rm h},0,b} - \xi_{{\rm r},0,b}\xi_{{\rm h},0,a} \right] 
\label{eq_Kab}
\end{align}
\end{linenomath*}
Eq. \ref{eq_freq_pert} shows that if $|\omega_a-\omega_b|\gg\omega_{1,ab}$, then we obtain $\omega_\pm\in\left\{\omega_a,\omega_b\right\}$, i.e. we recover the eigenfrequencies of the non-degenerate case.

As shown in Appendix \ref{app_neardeg}, we can also obtain the values of $A$ and $B$ in Eq. \ref{eq_eigenfunction} for each component of both multiplets, i.e. we can estimate the perturbations to the mode eigenfunctions caused by near-degeneracy effects (for non-degenerate modes, changes in the eigenfunctions perturb the mode frequencies only at second order). 

\subsection{Test on stellar models}

We tested the effects of near-degeneracy on the rotational multiplets using stellar models. For this purpose, we used the best-fit stellar model obtained for KIC7341231 by \cite{deheuvels12} (model B). This model was computed with the evolutionary code \cesam\ (\citealt{cesam}) and optimized to match the observed frequencies of the $l=1$ mixed modes for this star. We refer the reader to \cite{deheuvels12} for information about the input physics of this model and the optimization procedure that was adopted. One interesting feature of this model is that it also roughly reproduces the frequencies of the two $l=2$ mixed modes that are undergoing an avoided crossing between 380 and 384 $\mu$Hz (even though these modes were not included in the fitting procedure followed by \citealt{deheuvels12}). This model can therefore be used to investigate the origin of the observed asymmetries in these multiplets. 

We also had to assume a rotation profile for the star. For flexibility reasons, we chose an analytic rotation profile of the type
\begin{linenomath*}
\begin{equation}
\Omega(r) = \frac{\Omega_{\rm e}+\Omega_{\rm c}}{2} + \frac{\Omega_{\rm e}-\Omega_{\rm c}}{2} \tanh\left(\frac{r-r_0}{d}\right)
\label{eq_tanh}
\end{equation}
\end{linenomath*}
where $\Omega_{\rm c}$ and $\Omega_{\rm e}$ correspond to the core and surface rotation, respectively. We here took $\Omega_{\rm c}/2\pi=710$ nHz and $\Omega_{\rm e}/2\pi=150$ nHz (i.e. a core-envelope contrast of $\sim5$), as per the results obtained by \cite{deheuvels12} for KIC7341231. The influence of the level of radial differential rotation is studied in Sect. \ref{sect_asym_diffrot}. We assumed a smooth transition between the core and surface rotation rates ($d=0.06$) located inside the evanescent region ($r_0=0.2$). 

\begin{figure}
\begin{center}
\includegraphics[width=9cm]{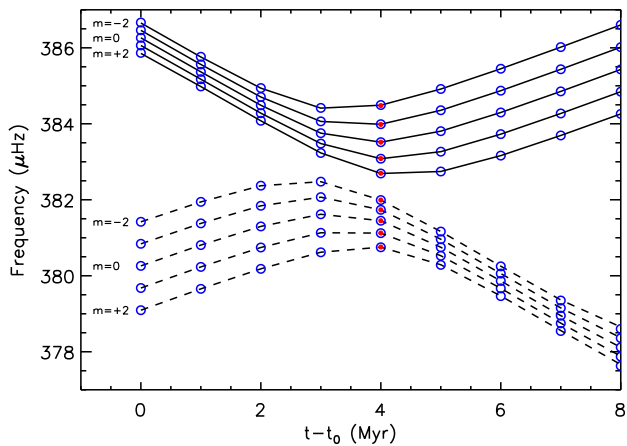}
\end{center}
\caption{Variations in the frequencies of two $l=2$ rotational multiplets during an avoided crossing. The sequence of models B0 through B8 was used, assuming that the rotation profile given by Eq. \ref{eq_tanh} with a core-envelope contrast of five. The frequencies of rotational multiplets were computed using first-order pertubative approach including near-degeneracy effects (open blue circles), and using the non-perturbative oscillation code \textsc{ACOR} (solid and dashed black curves). The filled red circles correspond to the frequencies obtained with the approximate Eq. \ref{eq_pert_omce} for model B4.
\label{fig_evol_multiplet_ACOR}}
\end{figure}

\begin{figure}
\begin{center}
\includegraphics[width=9cm]{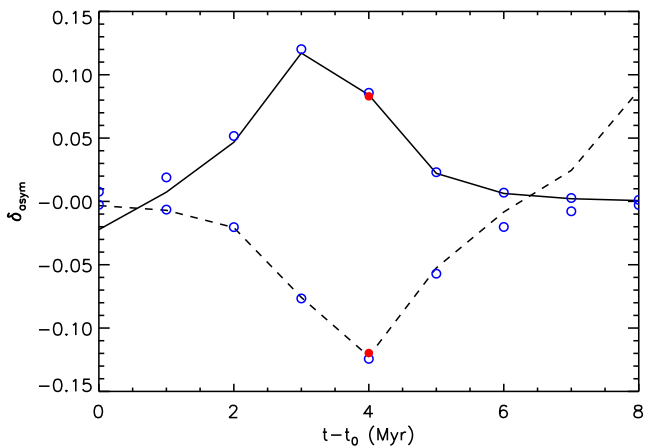}
\end{center}
\caption{Variations in the asymmetry of two $l=2$ rotational multiplets during an avoided crossing, computed using Eq. \ref{eq_asym} for $m=\pm2$ components. The symbols are identical to those of Fig. \ref{fig_evol_multiplet_ACOR}.
\label{fig_asym_multiplet}}
\end{figure}

To study the effects of near-degeneracy on the mode frequencies during the avoided crossing between the two $l=2$ multiplets, we recomputed the evolution of our best-fit model of KIC734123, but stopped the evolution at a time $t_0$ just before the avoided crossing (the corresponding model is hereafter referred to as model B0). We then resumed the evolution with small time steps (1 Myr) in order to span the entire avoided crossing. This sequence of models is labeled B0 through B8 and referred to as such in subsequent discussions. For each model in the sequence, we computed the unperturbed mode frequencies ($\omega_{0,a}$ and $\omega_{0,b}$) and eigenfunctions ($\vect{\xi}_{0,a}$ and $\vect{\xi}_{0,b}$) of the two $l=2$ modes that are in avoided crossing. We then calculated $\omega_{1,a}$ and $\omega_{1,b}$ using Eq. \ref{eq_om1}, and $\omega_{1,ab}$ using Eq. \ref{eq_om1ab}. The perturbed mode frequencies were then obtained using Eq. \ref{eq_freq_pert}. The variations in the mode frequencies for our sequence of models are plotted as a function of time in Fig. \ref{fig_evol_multiplet_ACOR} (blue open circles). Note that the frequencies of $m=0$ modes in fact correspond to the frequencies of unperturbed modes. Before the avoided crossing, the lowest-frequency multiplet has a g-mode behavior and the eigenfrequencies of its components thus increase owing to the core contraction. The highest-frequency multiplet behaves as a p mode and the frequencies of its components are decreasing owing to the increase of the stellar radius. When these two multiplets have comparable frequencies, the modes become mixed and the multiplets progressively exchange natures.

We found that the frequencies of the two $l=2$ multiplets do indeed show significant levels of asymmetry during the avoided crossing. We quantified these asymmetries by using Eq. \ref{eq_asym} for $m=\pm2$ components in the same way as was done with the observations. The results are shown in Fig. \ref{fig_asym_multiplet}. As expected, the asymmetries of the multiplets are maximal when the modes that are bumping each other are at their smallest frequency difference, i.e., when the modes are the most mixed. Away from the avoided crossing, the asymmetries decrease and eventually become negligible. 

The asymmetries obtained with our first-order perturbative approach including near-degeneracy effects show encouraging similarities with the ones observed in the spectrum of KIC7341231 using \kepler\ data. Indeed, the higher-frequency mode is found to have a positive $\delta_{\rm asym}$ during the avoided crossing, while the lower-frequency multiplet has a negative $\delta_{\rm asym}$. This is in agreement with the observed asymmetries (see Sect. \ref{sect_obs}). The orders of magnitude of the asymmetries are also comparable, although the ones found here are smaller. This difference is discussed in detail in Sect. \ref{sect_diffrot}.

\subsection{Mode eigenfunctions}

\begin{figure}
\begin{center}
\includegraphics[width=9cm]{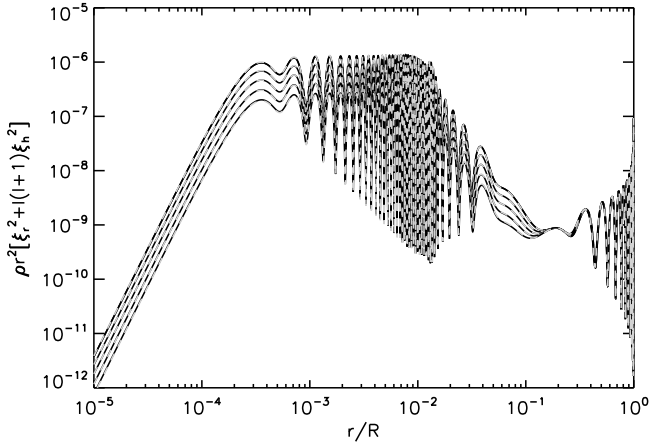}
\end{center}
\caption{Integrand of the mode inertia $\rho r^2 \left[\xi_{\rm r}^2+l(l+1)\xi_{\rm h}^2\right]$ for the $m$-components of the higher-frequency $l=2$ multiplet at the time during the avoided crossing where it shows the strongest asymmetry ($t=t_0+3$ Myr). The radial and horizontal displacements ($\xi_{\rm r}$ and $\xi_{\rm h}$) were computed using a perturbative approach as explained in Sect. \ref{sect_pert} (grey dashed lines) and with the code \textsc{ACOR} (solid black lines). 
\label{fig_fctpp}}
\end{figure}

As mentioned in Sect. \ref{sect_neardeg}, we were also able to estimate the eigenfunctions of the components of both multiplets for our sequence of models. These eigenfunctions can provide valuable insight about the trapping of each component during the avoided crossing. In Fig. \ref{fig_fctpp}, we show the integrand of the mode inertia $\rho r^2 \left[\xi_{\rm r}^2+l(l+1)\xi_{\rm h}^2\right]$ for all the $m$-components of the higher-frequency $l=2$ multiplet at the time during the avoided crossing when the asymmetry is the strongest ($t=t_0+3$ Myr). The radial displacements were normalized to their surface value, so that the contribution from the p-mode cavity is nearly identical for all the components. As can be seen in Fig. \ref{fig_fctpp}, the contribution of the core to the mode inertia varies by a factor of $\sim 10$ between the $m=-2$ and the $m=+2$ components. This shows that during the avoided crossing, the $m$-components of an $l=2$ multiplet are trapped differently inside the star. This sheds new light on the asymmetries of multiplets. Indeed, within a multiplet, the components that have a more g-like behavior mainly probe the core rotation, while the p-like components are more sensitive to the envelope rotation. This suggests that there is a link between the internal rotation of the star and the asymmetries in the rotational multiplets of mixed modes, which is addressed in the following section.

\subsection{Link between multiplet asymmetry and differential rotation \label{sect_asym_diffrot}}

The link between differential rotation and multiplet asymmetry can be simply understood based on Fig. \ref{fig_evol_multiplet_ACOR}. This figure shows that the avoided crossings between the $m$-components of the modes occur at slightly different epochs. The $m=+2$ modes are the first to bump with each other, followed by the $m=+1$ modes, and so on. The  time-delay between the avoided crossings of the different $m$-components is caused by the difference between the splittings of g-modes and those of p-modes. For the rotation profile that we assumed here, the splittings of g modes are larger than those of p modes. As a result, the $m=+2$ modes are the first ones to become close enough to each other for the coupling between p- and g-modes to start modifying their frequencies, as can be seen in Fig. \ref{fig_evol_multiplet_ACOR}. The link with multiplet asymmetries is clear. For instance, at $t-t_0=3$ Myr, in the lower-frequency multiplet, the $m<0$ components are still behaving as g modes, while the $m>0$ components start to resemble p modes because they are already coupling with the components of the higher-frequency mode. The splittings of the $m<0$ components are therefore larger than those of the $m>0$ components, thus producing a negative asymmetry (see Fig. \ref{fig_asym_multiplet}). Conversely, in the higher-frequency multiplet at that same time, the $m<0$ components are still p-like, while the $m>0$ components have already taken a g-like behavior. This produces a positive asymmetry, as confirmed by Fig. \ref{fig_asym_multiplet}.

\begin{figure}
\begin{center}
\includegraphics[width=9cm]{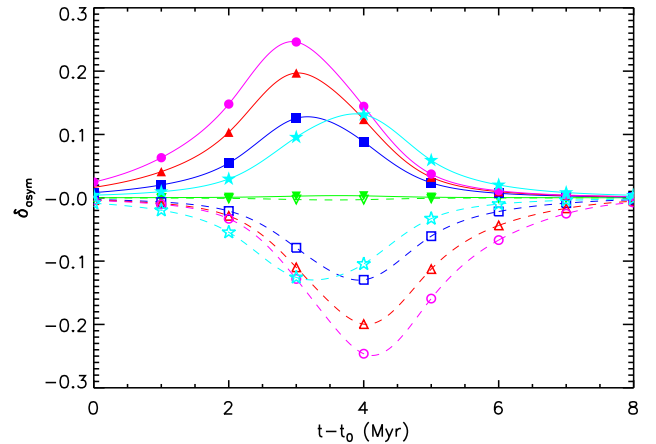}
\end{center}
\caption{Degree of asymmetry $\delta_{\rm asym}$ of the two $l=2$ multiplets during the avoided crossing. Different rotation profiles were considered, with core-envelope contrasts ($\Omega_{\rm c}/\Omega_{\rm e}$) of 20 (purple circles), 10 (red triangle), 5 (blue squares), 1 (green downward-triangle), and 0.5 (cyan stars). The higher-frequency multiplet is represented by filled symbols, and the lower-frequency multiplet by empty symbols. To guide the eye, we overplotted cubic spline interpolations $\delta_{\rm asym}(t-t_0)$ for each rotation profile --- solid (resp. dashed) lines for higher-frequency (resp. lower-frequency) multiplet.
\label{fig_asym_diffrot}}
\end{figure}

If this interpretation is correct, we expect the asymmetries to increase as the degree of differential rotation increases. Conversely, we expect the asymmetries to vanish when the p modes and the g modes have identical rotational splittings. For $l=2$ modes, this occurs when $\omc=6/5 \ome$, i.e., when the rotation is nearly solid-body like\footnote{Indeed, one can show that the rotational splitting of pure g-modes is approximately equal to $ \left( \frac{L^2-1}{L^2} \right) \omc$, while the splitting of pure p modes is approximately $\ome$}. When the envelope is spinning faster than the core, we expect the asymmetries to increase again as the envelope-to-core ratio increases. To check this, we modified the surface rotation $\Omega_{\rm e}$ in the input rotation profile given by Eq. \ref{eq_tanh} to produce core-envelope contrasts of 20, 10, 5, 1, and 0.5. For each of these rotation profiles, the mode frequencies were computed as before using Eq. \ref{eq_freq_pert} over the course of the avoided crossing. The asymmetries were then calculated using Eq. \ref{eq_asym}. We plotted the degree of asymmetry of both multiplets as a function of time in Fig. \ref{fig_asym_diffrot}. To guide the eye in this figure, we performed cubic spline interpolation of $\delta_{\rm asym}(t-t_0)$ for each rotation profile. When the core spins faster than the envelope, we observe that the asymmetry of the multiplets increases as the core-to-envelope rotation ratio increases, as predicted above. The asymmetries are at smallest in the case of a solid-body rotation. When the envelope spins faster than the core, the asymmetries increase again as the envelope-to-core rotation ratio increases, as we had foreseen. This clearly confirms that differential rotation is indeed the cause of multiplet asymmetries during avoided crossings. This opens the interesting possibility of using $l=2$ multiplets to probe the internal rotation of red giants. 

\subsection{Link between multiplet asymmetry and mode trapping \label{sect_asym_trapping}}

Fig. \ref{fig_asym_diffrot} also shows that the development of the asymmetry is not simultaneous for both multiplets during the avoided crossing.

When $\omc>6/5\ome$, the higher-frequency multiplet reaches its maximum asymmetry before the lower-frequency multiplet does (in terms of the absolute value). This can be understood as follows. Before the avoided crossing, the higher-frequency multiplet has a p-like behavior. At the beginning of the avoided crossing, the $m=+2$ component becomes more g-like and its splitting increases because of the fast core rotation. As a result, the numerator of $\delta_{\rm asym}$ in Eq. \ref{eq_asym} increases. Since the other components still have a p-like behavior, their rotational splittings are sensitive mainly to the envelope rotation and thus the denominator of $\delta_{\rm asym}$ remains small, leading to a large degree of asymmetry. On the other hand, the lower-frequency multiplet initially has a g-like behavior. When the avoided crossing begins, its $m=+2$ component takes on a p-like character and its splitting decreases because of the slow envelope rotation. The numerator of $\delta_{\rm asym}$ becomes increasingly negative but since the other components are still g-dominated, their splittings are sensitive to the core rotation and thus the denominator of $\delta_{\rm asym}$ remains large. As a result, $|\delta_{\rm asym}|$ is initially smaller than for the higher-frequency multiplet. Near the end of the avoided crossing, the situation is reversed and $|\delta_{\rm asym}|$ is larger for the lower-frequency multiplet. It is important to point this behavior out because it shows that the asymmetries of rotational multiplets carry information about how far the star is in the avoided crossing process, and thus on the trapping of the two $l=2$ modes. We make use of this property in Sect. \ref{sect_diffrot} when developing a procedure to infer core and envelope rotation rates from the frequencies of $l=2$ mixed modes. 

When $\omc<6/5\ome$, the splittings of $l=2$ p modes are larger than those of g modes, and the evolution of multiplet asymmetries  during the avoided crossing is the opposite of the previous case. We can indeed see in Fig. \ref{fig_asym_diffrot} that for $\omc/\ome=0.5$, the lower-frequency multiplet reaches its maximum of asymmetry (in absolute value) earlier than the higher-frequency multiplet. 

\section{Non-perturbative approach \label{sect_acor}}

Before tackling the question of probing the internal rotation of giants using $l=2$ rotational multiplets (see Sect. \ref{sect_diffrot}), we tested the validity of the first-order pertubative approach that includes near-degeneracy effects developed in Sect. \ref{sect_pert}. For this purpose, we used the oscillation code \textsc{ACOR} (Adiabatic Computations of Oscillations including Rotation, \citealt{ouazzani12}), which is one of only two codes that take into account the effects of rotation in a non-perturbative manner (the other being the Two-dimensional Oscillation Programme, a.k.a. \textsc{TOP}, developed by \citealt{reese06}). These codes were in fact designed to compute the pulsations of rapidly-rotating and therefore significantly distorted stars. 

\textsc{ACOR} solves the equations of hydrodynamics perturbed by Eulerian fluctuations. Rotation is included by taking into account both the effects of the centrifugal distortion and those of the Coriolis acceleration. The numerical method is based on a spectral multi-domain method that expands the angular dependence of pulsation modes on spherical-harmonic series. The radial differentiation is done by means of a sophisticated finite differences method which is accurate up to fifth order in terms of the radial resolution (\citealt{losc}). The code has been validated by comparison with the results of TOP (\citealt{reese06}) for polytropic models and the agreement between the two codes was found to be excellent (\citealt{ouazzani12}). 

We used \textsc{ACOR} to compute the mode frequencies for the sequence of models B0 through B8, which follows the avoided crossing between the two $l=2$ multiplets. We used the same rotation profile as in Sect. \ref{sect_pert} (Eq. \ref{eq_tanh} with a core-envelope contrast of five). The results are overplotted in Fig. \ref{fig_evol_multiplet_ACOR} (black solid and dashed lines). The agreement with the frequencies obtained with the first-order perturbative approach is excellent. 

We also calculated multiplet asymmetries $\delta_{\rm asym}$ by inserting \textsc{ACOR} mode frequencies into Eq. \ref{eq_asym}. As can be seen in Fig. \ref{fig_asym_multiplet}, the results match the asymmetries obtained with first-order perturbed frequencies in the vicinity of the avoided crossing (2 Myr $<t-t_0<$ 5 Myr) very well. Outside this range, differences appear between the two approaches. For $t-t_0\geqslant6$ Myr, the asymmetry of the lowest-frequency multiplet increases again, as can be seen in Fig. \ref{fig_asym_multiplet}. At this point, this mode still behaves mainly as a p mode, but it is already in the process of bumping with a gravity-dominated mode at lower frequency. Similarly, at $t=t_0$, the asymmetry of the largest-frequency multiplet is negative, which is caused by the bumping with another gravity-dominated mode at larger frequency. In comparison, the asymmetries predicted by the first-order perturbative approach vanish outside the avoided crossing because the two modes under consideration are no longer near-degenerate and adjacent $l=2$ modes were not included. This discrepancy could be resolved by including the two $l=2$ adjacent modes in the first-order perturbative approach; however, this is not necessary since $l=2$ mixed modes can be detected only when they have the most mixed character, i.e., only in the close vicinity of the maximum of an avoided crossing. 

Finally, we also computed the eigenfunctions of all components of the two multiplets along the sequence of models using \textsc{ACOR}. The radial and horizontal displacements were used to compute the integrand of the mode inertia for the higher-frequency $l=2$ multiplet at $t=t_0+3$ Myr, and the results were overplotted in Fig. \ref{fig_fctpp}. The agreement with the eigenfunctions obtained with the first-order perturbative approach is once again excellent.

These comparisons clearly validate the use of a first-order perturbative approach taking near-degeneracy into account to estimate the effects of rotation on the frequencies of mixed modes during avoided crossings. In particular, they confirm that second-order effects of rotation on the mode frequencies can indeed be safely neglected here.


\section{Probing internal rotation using $l=2$ multiplets \label{sect_diffrot}}

We have shown in Sect. \ref{sect_pert} that the asymmetry of $l=2$ multiplets depends directly on the amount of differential rotation. In this section, we investigate whether the frequencies of $l=2$ asymmetric multiplets can be used to measure mean core rotation $\omc$ and envelope rotation $\ome$ in red giants. For this purpose, the perturbative expression of the frequency shifts caused by rotation given by Eq. \ref{eq_freq_pert} is convenient because it opens up the possibility for applying inversion methods to recover the internal rotation profile. 

\subsection{A simplified expression for first-order perturbation of mode frequencies}

\begin{figure}
\begin{center}
\includegraphics[width=9cm]{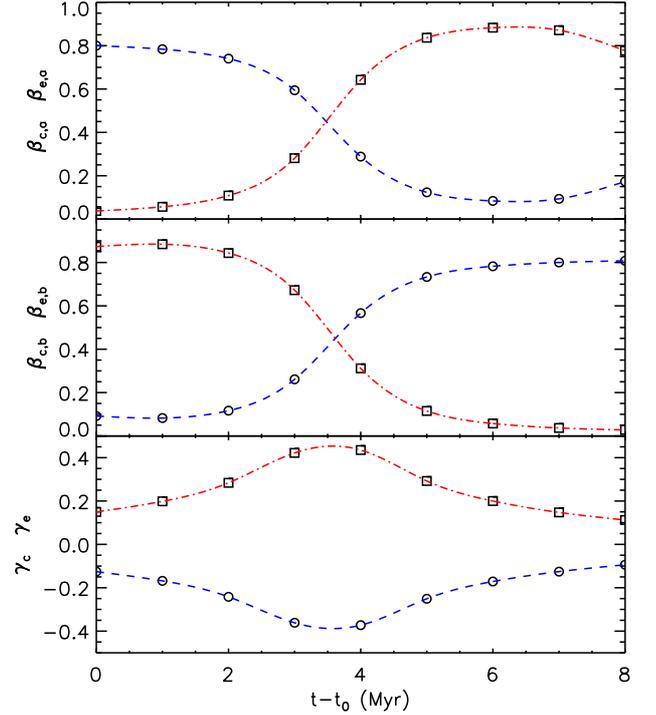}
\end{center}
\caption{Variations in the parameters $\beta_{\rm c}$, $\beta_{\rm e}$, $\gamma_{\rm c}$, and $\gamma_{\rm e}$ (defined in Eq. \ref{eq_betac}, \ref{eq_betae}, \ref{eq_gammac}, and \ref{eq_gammae}) along the course of the avoided crossing between the two $l=2$ multiplets for our reference sequence of models B0 through B8. Mode $a$ (resp. $b$) correspond to the lower- (resp. higher-) frequency multiplet. Core quantities are represented by open circles and envelope quantities by open squares. Interpolations using cubic splines are overplotted (blue dashed lines for core parameters and red dot-dash lines for envelope parameters).
\label{fig_beta_gamma}}
\end{figure}

We separated the first-order perturbation of mode eigenfrequencies into contributions from the core and the envelope. For instance, we write for mode $a$ (which we here considered to correspond to the lower-frequency multiplet)
\begin{linenomath*}
\begin{equation}
\omega_{1,a} = m \left( \beta_{{\rm c},a}\omc + \beta_{{\rm e},a}\ome \right)
\end{equation}
\end{linenomath*}
where
\begin{linenomath*}
\begin{align}
\beta_{{\rm c},a} & \equiv \int_0^{r_{\rm c}} K_a(r) \,\hbox{d}r \label{eq_betac} \\
\beta_{{\rm e},a} & \equiv \int_{r_{\rm c}}^R K_a(r) \,\hbox{d}r  \label{eq_betae}
\end{align}
\end{linenomath*}
and
\begin{linenomath*}
\begin{align}
\omc \equiv \frac{\int_0^{r_{\rm c}} K_a(r) \Omega(r) \,\hbox{d}r }{\int_0^{r_{\rm c}} K_a(r) \,\hbox{d}r} \label{eq_omc} \\
\ome \equiv \frac{\int_{r_{\rm c}}^R K_a(r) \Omega(r) \,\hbox{d}r }{ \int_{r_{\rm c}}^R K_a(r) \,\hbox{d}r }\label{eq_ome}
\end{align}
\end{linenomath*}
The kernel $K_a(r)$ is obtained from Eq. \ref{eq_K1}. The radius $r_{\rm c}$ was defined as the outer turning point of the gravity waves ($N(r_{\rm c})=\omega_{0,a}$, where $N(r)$ is the \vaisala\ frequency), so that $\omc$ is an average of the rotation rate over the g-mode cavity. Correspondingly, $\ome$ is an average of the angular velocity in the layers above $r_{\rm c}$. It is essentially a measure of the mean rotation rate in the p-mode cavity because the evanescent zone (region between $r_{\rm c}$ and the inner turning point of the p-mode cavity) contributes very little to this average. We also introduced $\beta_{{\rm c},b}$ and $\beta_{{\rm e},b}$ for mode $b$ (corresponding to the higher-frequency multiplet) with expressions analogous to Eq. \ref{eq_betac} and \ref{eq_betae}. We note that  $r_{\rm c}$ varies little with mode frequency because of the sharp decrease of the \vaisala\ frequency at the edges of the g-mode cavity. We thus took the same value for both modes. Likewise, the values of $\omc$ and $\ome$ are very similar whether we use the kernel of mode $a$ or that of mode $b$ in Eq. \ref{eq_omc} and \ref{eq_ome}. In the following discussion, we will thus refer to these quantities without mentioning which kernel was used to compute them. The values of $\beta_{\rm c}$ and $\beta_{\rm e}$ were computed for modes $a$ and $b$ for our sequence of models spanning the avoided crossing between the two $l=2$ modes. The results are shown in Fig. \ref{fig_beta_gamma}. One can clearly see that mode $a$ (lower-frequency multiplet) is initially sensitive to core rotation (g-dominated) and its sensitivity gradually shifts to the envelope (p-dominated), while mode $b$ (higher-frequency multiiplet) has the opposite behavior.

The coupling term $\omega_{1,ab}$ was also rewritten as
\begin{linenomath*}
\begin{equation}
\omega_{1,ab} = m \left( \gamma_{\rm c}\omc + \gamma_{\rm e}\ome \right)
\label{eq_omega1ab}
\end{equation}
\end{linenomath*}
where
\begin{linenomath*}
\begin{align}
\gamma_{\rm c} & \equiv \int_0^{r_{\rm c}} K_{ab}(r) \,\hbox{d}r \label{eq_gammac} \\
\gamma_{\rm e} & \equiv \int_{r_{\rm c}}^R K_{ab}(r) \,\hbox{d}r  \label{eq_gammae}
\end{align}
\end{linenomath*}
and the function $K_{ab}(r)$ is defined by Eq. \ref{eq_om1ab}. The values of $\gamma_{\rm c}$ and $\gamma_{\rm e}$ during the avoided crossing are shown in Fig. \ref{fig_beta_gamma}. As described in Appendix \ref{app_omega1ab}, a simplified expression can be obtained for $\omega_{1,ab}$, which shows that the coupling term vanishes in the particular case of a rotation profile for which the splittings of pure g modes are identical to those of pure p modes. In this case, we recover the eigenfrequencies of the non-degenerate case (see Eq. \ref{eq_freq_pert}) and the multiplets remain symmetric, in agreement with what was found in Sect. \ref{sect_asym_diffrot}.

As a result, the first-order perturbed mode frequencies can be expressed as
\begin{linenomath*}
\begin{align}
\omega_{\pm} &   = \bar{\omega}_0+m\left(\bar{\beta}_{\rm c}\omc + \bar{\beta}_{\rm e}\ome \right) \pm \frac{1}{2} \left\{ 4m^2(\gamma_{\rm c}\omc+\gamma_{\rm e}\ome)^2 \right. \nonumber \\
& \left. + \left[\delta\omega_0 + m(\delta\beta_{\rm c}\omc+\delta\beta_{\rm e}\ome)\right]^2 \right\}^{1/2} \label{eq_pert_omce}
\end{align}
\end{linenomath*}
where we have defined
\begin{linenomath*}
\begin{align}
\bar{\omega}_0 \equiv \frac{\omega_{0,a}+\omega_{0,b}}{2} \;\; & ; \;\; \delta\omega_0 \equiv \omega_{0,a}-\omega_{0,b} \\
\bar{\beta_{\rm c}} \equiv \frac{\beta_{{\rm c},a}+\beta_{{\rm c},b}}{2} \;\; & ; \;\;  \delta\beta_{\rm c} \equiv \beta_{{\rm c},a}-\beta_{{\rm c},b} 
\end{align}
\end{linenomath*}
and similar expressions hold for $\bar{\beta_{\rm e}}$ and $\delta\beta_{\rm e}$ as well. 

We thus obtained approximate expressions for the perturbed mode frequencies involving only four parameters ($\omc$, $\ome$, $\omega_{0,a}$, and $\omega_{0,b}$) and the rotational kernels of the chosen reference model. This gives us the possibility of estimating the mean core and envelope rotation using $l=2$ modes by fitting Eq. \ref{eq_pert_omce} to the observed mode frequencies. We note that $\beta_{\rm c}$ and $\beta_{\rm e}$ can in fact be approximately estimated from seismic observables without the need of a reference stellar model, as was shown by \cite{goupil13} (see also \citealt{mosser15}). If approximate expressions relating $\gamma_{\rm c}$ and $\gamma_{\rm e}$ to observable quantities can be derived, then Eq. \ref{eq_pert_omce} could be used to estimate $\omc$ and $\ome$ directly from the observed mode frequencies. This is however out of the scope of the present work, and instead we compute $\beta_{\rm c}$, $\beta_{\rm e}$, $\gamma_{\rm c}$, and $\gamma_{\rm e}$ from reference stellar models in this study.

\subsection{Preliminary test}

To test this approach, we used the eigenfrequencies obtained with \textsc{ACOR} for a model in the close neighborhood of the avoided crossing between the $l=2$ multiplets (model B4) as artificial data, assuming the same rotation profile as in Sect. \ref{sect_pert} (Eq. \ref{eq_tanh} with a core-envelope contrast of five). In this preliminary test, we assumed that all the components of the multiplets are detected (ten frequencies) and that the frequencies are free of noise. We used the same model (model B4) as the reference model to perform inversions of $\omc$ and $\ome$. 
We used the non-rotating version of \textsc{ACOR} to compute the unperturbed eigenfunctions of model B4, which were then inserted into Eqs. \ref{eq_betac}, \ref{eq_betae}, \ref{eq_gammac}, and \ref{eq_gammae} to obtain the values of $\beta_{\rm c}$, $\beta_{\rm e}$ for both modes, as well as $\gamma_{\rm c}$, $\gamma_{\rm e}$. We then performed an iterative fit of Eq. \ref{eq_pert_omce} to the eigenfrequencies obtained with \textsc{ACOR}, using $\omc$, $\ome$, $\omega_{0,a}$, and $\omega_{0,b}$ as free parameters of the fit. The optimal values obtained for the core and envelope rotation rates ($\omtot_{\rm c,fit}/2\pi=716$ nHz and $\omtot_{\rm e,fit}/2\pi=154$ nHz) are very close to the true values of these quantities computed using Eq. \ref{eq_omc} and \ref{eq_ome} for our reference rotation profile ($\omtot_{\rm c,mod}/2\pi=710$ nHz, and $\omtot_{\rm e,mod}/2\pi=154$ nHz). The frequencies obtained using the best-fit parameters and Eq. \ref{eq_pert_omce} are overplotted in Fig. \ref{fig_evol_multiplet_ACOR} (red filled circles at $t=t_0+4$ Myr). The agreement with the frequencies computed with \textsc{ACOR} is better than 10 nHz, i.e., below the precision to which rotational splittings can be measured with the four years of \kepler\ observations.

These results are encouraging, but they do not guarantee that we can recover $\omc$ and $\ome$ using the observed frequencies of $l=2$ multiplets. Indeed, in this preliminary test, we considered the ideal case where the reference model is ``perfect''. If we perform a similar fit, but use the values of $\beta_{\rm c}$, $\beta_{\rm e}$, $\gamma_{\rm c}$, and $\gamma_{\rm e}$ computed with a different model, (say for instance model B5), no satisfactory fit to the frequencies of \textsc{ACOR} can be found and the recovered values of $\omc$ and $\ome$ are incorrect. The reason for this is that the trapping of the modes differs between the chosen reference model and the model that was used to produce rotational multiplets. 

The need for a reference model that produces a correct trapping of the modes was already emphasized in works that performed rotation inversions using dipolar modes. However, in the latter case, a large number of $l=1$ modes can be detected, and these have a wide variety of natures ranging from g-dominated to p-dominated. \cite{deheuvels12} showed that all the non-rotating models that provides satisfactory matches to the frequencies of the $m=0$ components of the observed $l=1$ modes predict a similar trapping of the dipolar modes, quantified by the parameter $\zeta$ defined as the ratio between the mode inertia in the g-mode cavity and the total mode inertia, i.e.
\begin{linenomath*}
\begin{equation}
\zeta \equiv \frac{\int_0^{r_{\rm c}} \rho r^2 ( \xi_{\rm r}^2 + L^2\xi_{\rm h}^2 ) \hbox{d}r}{\int_0^R \rho r^2 ( \xi_{\rm r}^2 + L^2\xi_{\rm h}^2 ) \hbox{d}r}.
\label{eq_zeta}
\end{equation}
\end{linenomath*}
\cite{goupil13} showed that the trapping of $l=1$ modes can even be estimated reliably from the mode frequencies themselves, independently of stellar models, using the JWKB approximation. For $l=2$ modes, the situation seems much more complex. \cite{deheuvels12} had mentioned that the models that satisfactorily reproduce the frequencies of $l=1$ modes predict different mode trappings for the $l=2$ mixed modes, which means that these models cannot be directly used to perform inversions using $l=2$ modes. Besides, the coupling between the p-mode and g-mode cavities is weaker for $l=2$ modes than for $l=1$ modes because the evanescent region that separates them is wider. We can thus detect $l=2$ modes only when their frequencies are very close to those of pure p modes. Outside these frequency ranges, the $l=2$ modes have inertias that are too large for the modes to have detectable heights in the oscillation spectra, even with the longest \kepler\ datasets (see e.g. \citealt{grosjean14}). This makes it much harder to precisely estimate the asymptotic properties of $l=2$ pure g modes, which are required if we want to apply the method of \cite{goupil13} to $l=2$ modes.


\subsection{New fitting procedure}

In Sect. \ref{sect_asym_diffrot}, we have shown that the asymmetries of the rotational multiplets carry information not only about the internal rotation, but also about how far the star is in the avoided crossing process, and thus on the trapping of the $l=2$ modes. This could yield a solution to the problem mentioned above. We explored the possibility of also recovering the trapping of the $l=2$ modes (expressed by the $\zeta$ parameter) when fitting Eq. \ref{eq_pert_omce} to the data. The idea is to allow the age of the reference model to vary slightly in our fitting procedure by considering the time elapsed since the beginning of the avoided crossing, $\Delta t \equiv t-t_0$, as an additional free parameter. We note that because of the much weaker coupling, the timescale over which the trapping of $l=2$ modes is modified is small compared to the equivalent quantity for $l=1$ modes. As a result, the small age adjustment that is made in this fit has negligible impact on the trapping of $l=1$ modes in the model.  

With this new fitting procedure, the value of $\Delta t$ changes at each iteration, and in principle the rotational kernels need to be recomputed every time, which is numerically very costly. In fact the only quantities that are needed are $\beta_{\rm c}$, $\beta_{\rm e}$, $\gamma_{\rm c}$, and $\gamma_{\rm e}$, whose variations during the avoided crossing are quite smooth (see Fig. \ref{fig_beta_gamma}). We thus chose to perform cubic-spline interpolation of these quantities using our sequence of models B0 through B8. The results are overplotted in Fig. \ref{fig_beta_gamma}. For several values of $\Delta t$, we compared the interpolated values of $\beta_{\rm c}$, $\beta_{\rm e}$, $\gamma_{\rm c}$, and $\gamma_{\rm e}$ to the ones computed using the rotational kernels and found agreements at a level of 1\% or better. Inserting the interpolated parameters into Eq. \ref{eq_pert_omce}, this 1\% discrepancy translates into frequency differences at most of the order of 1 nHz, which is small compared to the precision to which the mode frequencies can be measured from the observations. We thus used interpolated values of $\beta_{\rm c}$, $\beta_{\rm e}$, $\gamma_{\rm c}$, and $\gamma_{\rm e}$ in the fits presented in the subsequent sections.

\avoir{For clarity, we briefly summarize the steps of the fitting procedure
\begin{enumerate}
\item Compute a reference model that satisfactorily reproduces the observational constraints of the star under study. This model should in particular reproduce the frequencies of the observed $l=1$ modes and roughly match the frequencies of the $l=2$ modes that are undergoing an avoided crossing.
\item Recompute the evolution of the reference model until a time $t_0$ just before the avoided crossing
\item Restart the evolution with small timesteps to span the entire avoided crossing. 
\item For each model of the sequence, compute the values of the $\beta$ and $\gamma$ parameters using Eq. \ref{eq_betac}, \ref{eq_betae}, \ref{eq_gammac}, and \ref{eq_gammae}.
\item Perform cubic spline interpolations of the variations in the $\beta$ and $\gamma$ parameters with time so that they can be estimated at any time during the avoided crossing
\item Perform an iterative fit of Eq. \ref{eq_pert_omce} to the observed frequencies of the component of the two $l=2$ multiplets, using $\omc$, $\ome$, $\omega_{0,a}$, and $\omega_{0,b}$ as free parameters. At each iteration of the fit, the $\beta$ and $\gamma$ parameters corresponding to the value of $\Delta t$ are calculated using the interpolations.
\end{enumerate}}

\begin{table*}
  \begin{center}
  \caption{Values of the parameters $\omc$, $\ome$, $\Delta t$, $\zeta_{0,a}$, and $\zeta_{0,b}$ obtained when fitting Eq. \ref{eq_pert_omce} to the mode frequencies of \textsc{ACOR} for models A, C, D, and E, assuming a rotation profile following Eq. \ref{eq_tanh} with a core rotation $\omc/2\pi=710$ nHz and core-envelope contrasts of 100, 20, 10, 5, and 2. For comparison, the right columns give the actual values of $\omc$, $\ome$, $\zeta_{0,a}$, and $\zeta_{0,b}$ computed with the input rotation profile and the eigenfunctions of the chosen model.
 \label{tab_test_model}}
\begin{tabular}{l c | c c c c c | c c c c} 
\hline \hline
\T Model id. & Rotation & \multicolumn{5}{c}{Fit} & \multicolumn{4}{c}{Model} \\
 & profile & $\omc/2\pi$ & $\ome/2\pi$ & $\Delta t$ & $\zeta_{0,a}$ & $\zeta_{0,b}$ & $\omc/2\pi$ & $\ome/2\pi$ & $\zeta_{0,a}$ & $\zeta_{0,b}$ \\
\B & & (nHz) & (nHz) & & & & (nHz) & (nHz) & & \\
\hline

\T Model A & 100 &  714.7 &  21.5 & 4.201 & 0.287 & 0.738 &  709.6 &  22.7 & 0.286 & 0.737 \\
 &  20 &  714.6 &  49.5 & 4.202 & 0.287 & 0.738 &  709.6 &  50.5 & 0.286 & 0.737 \\
 &  10 &  714.4 &  84.5 & 4.202 & 0.287 & 0.738 &  709.6 &  85.2 & 0.286 & 0.737 \\
 &   5 &  714.1 & 154.5 & 4.204 & 0.287 & 0.738 &  709.7 & 154.6 & 0.286 & 0.737 \\
\B &   2 &  713.2 & 364.4 & 4.214 & 0.283 & 0.743 &  709.8 & 362.9 & 0.286 & 0.737 \\
\T Model C & 100 &  720.8 &  18.1 & 3.824 & 0.407 & 0.619 &  709.6 &  23.6 & 0.406 & 0.621 \\
 &  20 &  720.3 &  46.3 & 3.824 & 0.407 & 0.619 &  709.6 &  51.4 & 0.406 & 0.621 \\
 &  10 &  719.8 &  81.5 & 3.824 & 0.407 & 0.619 &  709.6 &  86.0 & 0.406 & 0.621 \\
 &   5 &  718.6 & 151.8 & 3.825 & 0.407 & 0.619 &  709.7 & 155.4 & 0.406 & 0.621 \\
\B &   2 &  715.2 & 363.0 & 3.830 & 0.404 & 0.622 &  709.8 & 363.4 & 0.406 & 0.621 \\
\T Model D & 100 &  724.1 &  15.8 & 3.630 & 0.480 & 0.546 &  709.6 &  24.8 & 0.481 & 0.548 \\
 &  20 &  723.5 &  44.1 & 3.630 & 0.480 & 0.546 &  709.6 &  52.5 & 0.481 & 0.548 \\
 &  10 &  722.7 &  79.4 & 3.630 & 0.480 & 0.546 &  709.6 &  87.1 & 0.481 & 0.548 \\
 &   5 &  721.1 & 150.0 & 3.631 & 0.480 & 0.546 &  709.7 & 156.3 & 0.481 & 0.548 \\
\B &   2 &  716.3 & 362.0 & 3.633 & 0.480 & 0.546 &  709.8 & 363.9 & 0.481 & 0.548 \\
\T Model E & 100 &  723.2 &  17.2 & 3.685 & 0.458 & 0.568 &  709.6 &  24.1 & 0.460 & 0.570 \\
 &  20 &  722.6 &  45.4 & 3.685 & 0.458 & 0.568 &  709.6 &  51.8 & 0.460 & 0.570 \\
 &  10 &  721.9 &  80.6 & 3.685 & 0.458 & 0.568 &  709.6 &  86.4 & 0.460 & 0.570 \\
 &   5 &  720.4 & 151.0 & 3.686 & 0.458 & 0.568 &  709.7 & 155.7 & 0.460 & 0.570 \\
\B &   2 &  715.9 & 362.3 & 3.689 & 0.458 & 0.568 &  709.8 & 363.6 & 0.460 & 0.570 \\
\hline
\end{tabular}
\end{center}
\end{table*}

\subsection{Test of the method with noise-free artificial data \label{sect_simu_ideal} }

To test the fitting procedure described above, we again generated artificial data (frequencies of $l=2$ rotational multiplets in avoided crossing) using other models than model B. We conveniently used models A, C, D, and E obtained by \cite{deheuvels12}. All these models were found to reproduce the observed frequencies of the $m=0$ components of the observed $l=1$ mixed modes of KIC7341231 quite satisfactorily. They also match the observed atmospheric parameters of the star. The main difference between these models is the metallicity, which ranges from $-0.75$ dex (model A) to $-1.75$ dex (model E), reflecting the poorly constrained metallicity of the star. All these models were used by \cite{deheuvels12} to perform inversions of the rotation profile of KIC7341231, and were found to yield almost identical results. They all include an $l=2$ avoided crossing between two $l=2$ modes around 380 $\mu$Hz but they predict different mode trappings for these two $l=2$ mixed modes. For models A, C, D, and E, we computed the frequencies of the two $l=2$ rotational multiplets with \textsc{ACOR}, assuming the same rotation profile as before (Eq. \ref{eq_tanh} with a core rotation of 710 nHz) and we considered core-envelope contrasts of 100, 20, 10, 5, and 2. 

For each of these scenarios, we fitted Eq. \ref{eq_pert_omce} to the frequencies of \textsc{ACOR}, considering $\omc$, $\ome$, $\omega_{0,a}$, $\omega_{0,b}$, and $\Delta t$ as free parameters. As described above, for each iteration of the fit we interpolated through the values of $\beta_{\rm c}$, $\beta_{\rm e}$, $\gamma_{\rm c}$, and $\gamma_{\rm e}$ of the sequence of models B0 though B8. The inversions of $\omc$ and $\ome$ are thus performed using a different model than the ones used to generate the artificial data. The results are given in Table \ref{tab_test_model}. The first observation is that the recovered mean core and envelope rotation rates agree very well with the actual values computed with the input rotation profile and Eq. \ref{eq_omc} and \ref{eq_ome}, regardless of the chosen core-envelope ratio and the model that is used to generate artificial data.

The optimal value of $\Delta t$ can also be used to obtain an estimate of $\zeta_{0,a}$ and $\zeta_{0,b}$, which quantify the trapping of the $m=0$ components of the two $l=2$ multiplets. Indeed, we can compute $\zeta_{0,a}$ and $\zeta_{0,b}$ for the models B0 through B8 and perform cubic spline interpolations of these quantities in the same way as was done for the $\beta$ and $\gamma$ parameters. The values of $\zeta_{0,a}$ and $\zeta_{0,b}$ can then be obtained at the fitted time $\Delta t$. The corresponding values are given in Table \ref{tab_test_model}. They can be compared to the actual values of $\zeta_{0,a}$ and $\zeta_{0,b}$ computed by injecting the eigenfunctions of the model that was used to generate data into Eq. \ref{eq_zeta}. As can be seen in Table \ref{tab_test_model}, the agreement between the recovered values of $\zeta_{0,a}$ and $\zeta_{0,b}$ and the actual ones computed from the model is excellent. This confirms that we can indeed determine the trapping of the modes using the frequencies of two $l=2$ rotational multiplets in avoided crossing.

\subsection{Test of the method under realistic conditions \label{sect_simu_noise} }

Next, we tested our method with more realistic artificial data by including noise and by taking into account that the observed $l=2$ multiplets are generally incomplete. We performed Monte Carlo simulations using models A, C, D, and E and considered different rotation profiles. In the following, we show only the results obtained with model D (the other models give undistinguishable results), assuming a rotation profile with $\omc/2\pi=710$ nHz as before and a core-envelope contrast of 5. The influence of a change in this parameter is mentioned when relevant.

\subsubsection{Effects of adding noise on mode frequencies \label{sect_noise}}

\begin{figure}
\begin{center}
\includegraphics[width=9cm]{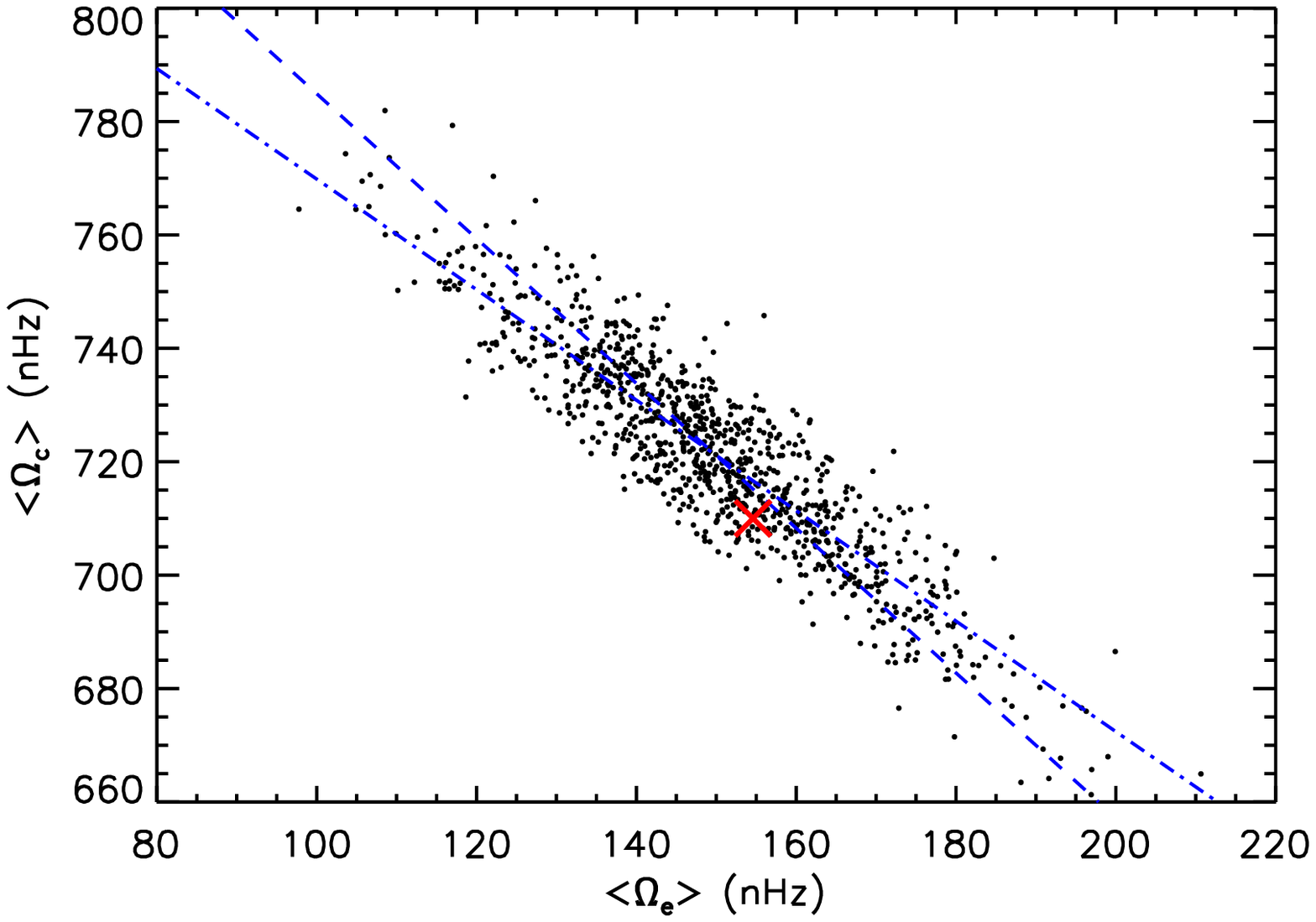}
\end{center}
\caption{Values of $\omc$ and $\ome$ obtained by fitting Eq. \ref{eq_pert_omce} to the mode frequencies of model D computed with \textsc{ACOR}, assuming a rotation profile with $\omc/2\pi=710$ nHz and $\omc/\ome=5$. A random noise with an rms of 15 nHz was added to the mode frequencies and 1000 iterations were performed. The quantity ${\beta_{\rm c}\omc+\beta_{\rm e}\ome}$ is conserved along the blue lines, where $\beta_{\rm c}$ and $\beta_{\rm e}$ correspond to mode $a$ (dashed line) or mode $b$ (dot-dashed line). The red cross indicates the actual values of $\omc$ and $\ome$ computed with Eq. \ref{eq_omc} and \ref{eq_ome}.
\label{fig_omc_ome}}
\end{figure}

We first assumed uncertainties of 15 nHz for the frequencies of the components of rotation multiplets. This value corresponds to the typical error bar resulting from the extraction of mode frequencies in the oscillation spectra of red giants observed during four years with \kepler\ (see for instance Table \ref{tab_otto_freq}). We added a realization of Gaussian noise with a standard deviation of 15 nHz to the mode frequencies computed with \textsc{ACOR}  and fitted Eq. \ref{eq_pert_omce} to the resulting set of frequencies. This operation was then repeated 1000 times. The values of $\omc$ and $\ome$ recovered from our Monte Carlo simulation for this case are shown in Fig. \ref{fig_omc_ome}. Using the distribution of the fitted parameters, we derived $\omc/2\pi=721\pm18$ nHz and $\ome/2\pi=150\pm16$ nHz, which is in close agreement with the actual values of $\omc$ and $\ome$ computed using Eq. \ref{eq_omc} and \ref{eq_ome} (see Table \ref{tab_test_model}). It is clear from Fig. \ref{fig_omc_ome} that even with noisy data one can efficiently recover the average core and envelope rotation rates. 

We repeated this procedure using different sets of initial parameters for the iterative fits. For instance, we tried starting from a solid-body rotation or with profiles such that $\ome>\omc$. We also varied the initial value of $\Delta t$. In all cases, the obtained results are undistinguishable from the ones mentioned above, which shows that the procedure has little dependence on the chosen initial parameters. We note however, that an ambiguity may arise in the determination of $\Delta t$. As was mentioned in Sect. \ref{sect_asym_diffrot}, a larger $|\delta_{\rm asym}|$ for the higher-frequency multiplet means that the star is either at the beginning of the avoided crossing with a core rotating faster than the envelope, or near the end of the avoided crossing with an envelope rotating faster the core (see Fig. \ref{fig_asym_diffrot}). This potential degeneracy was not found when using models A, C, D, or E to generate artificial data, but it arose when using model B3. When choosing $\omc>\ome$ in the initial parameters of the fit, we recovered the correct mean core and envelope rotation rates with $\Delta t=3$ Myr, as expected. However, when starting with $\ome>\omc$, the fit converged to a spurious solution with a fast rotating envelope near the end of the avoided crossing ($\omc/2\pi=194$ nHz and $\ome/2\pi=621$ nHz, with $\Delta t=5.1$ Myr). However, if this potential ambiguity arises, it should be easy to resolve, since the two scenarios provide very different splittings for dipolar modes. In all the red giants studied so far using \kepler\ data, the rotational splittings of $l=1$ modes safely ruled out the possibility of an envelope rotating faster than the core.

\begin{figure*}
\begin{center}
\includegraphics[width=9cm]{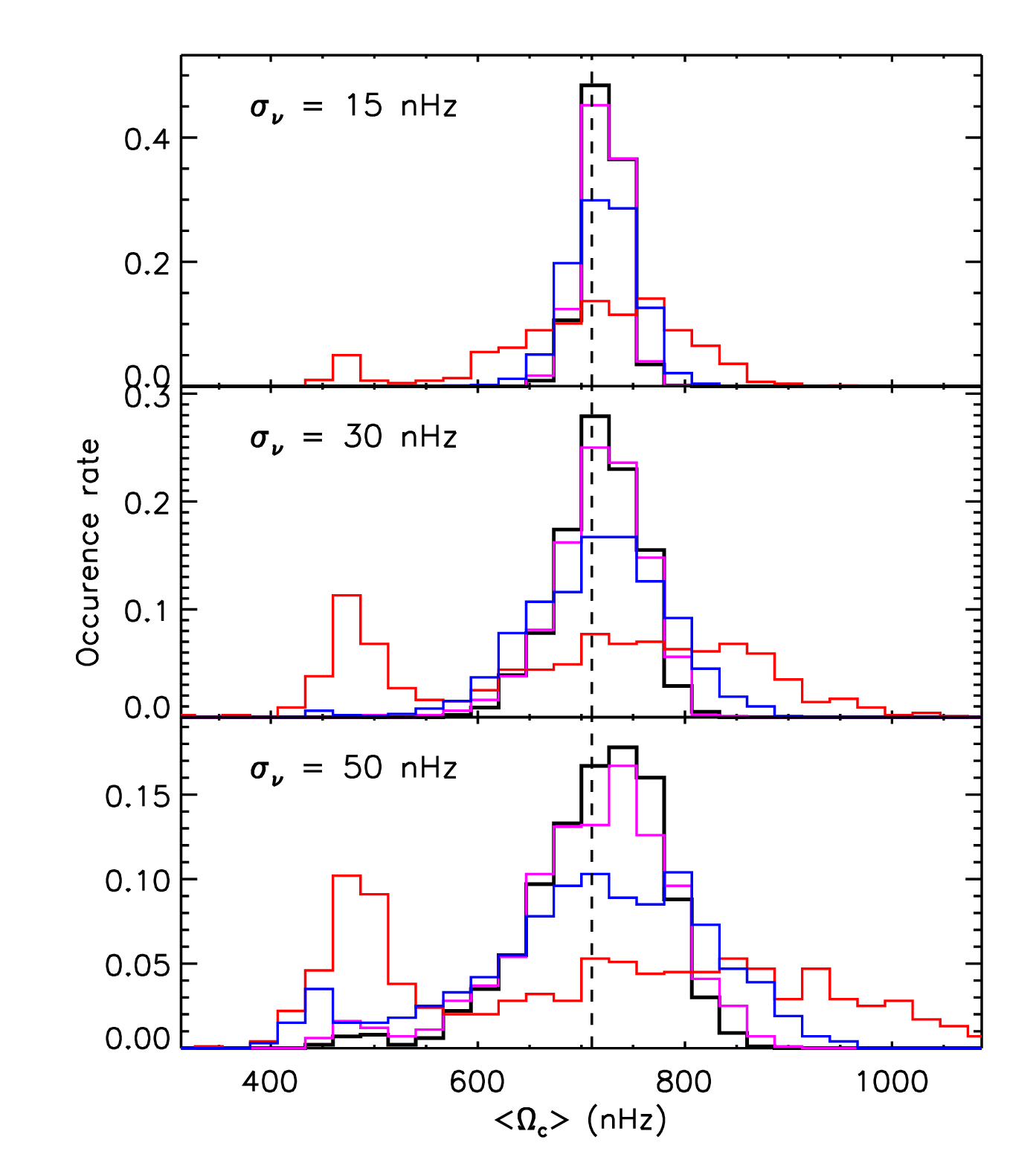}
\includegraphics[width=9cm]{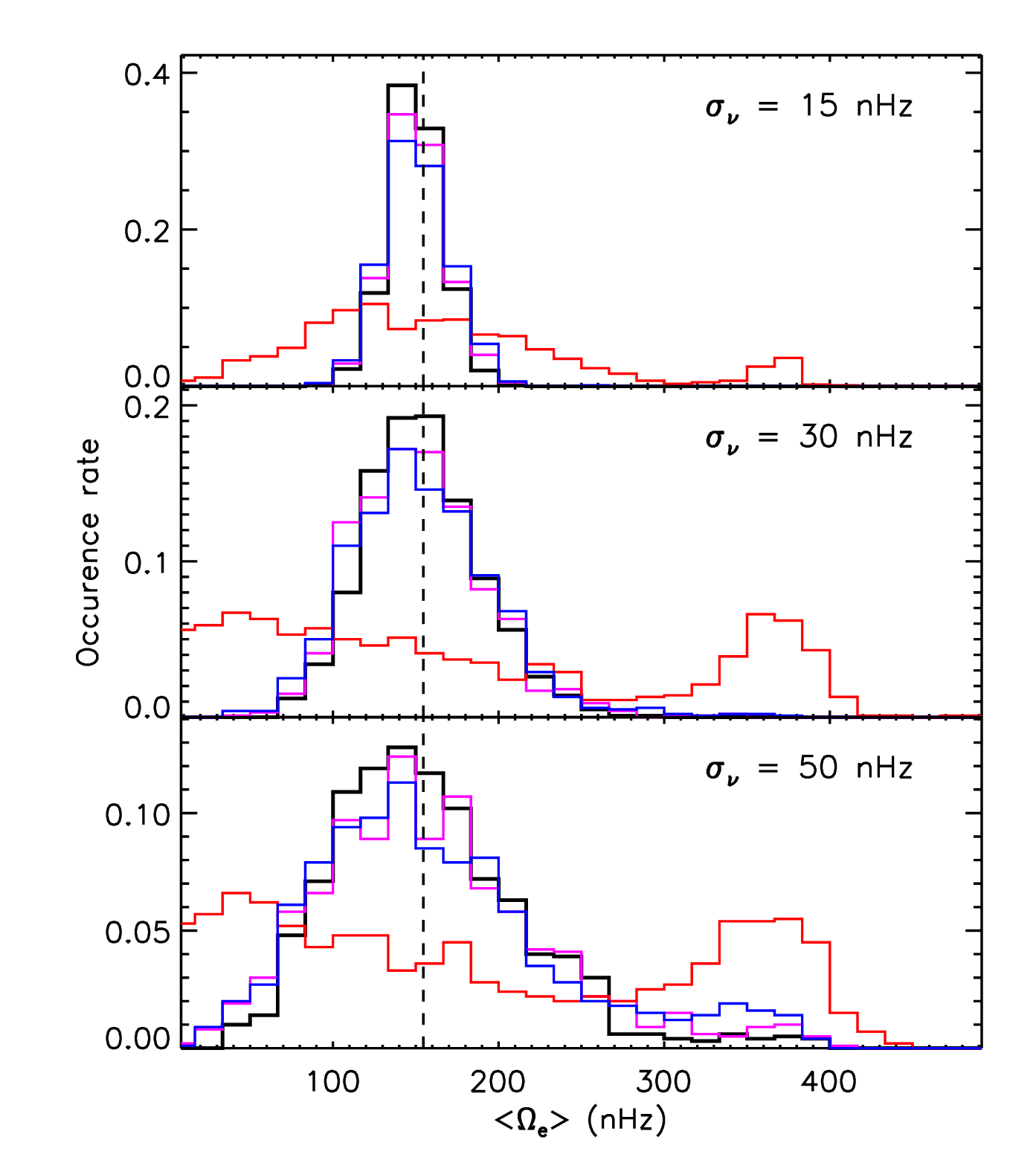}
\end{center}
\caption{Distribution of the values of $\omc$ (left) and $\ome$ (right) obtained by fitting Eq. \ref{eq_pert_omce} to the mode frequencies of \textsc{ACOR} for model D, assuming a rotation profile with $\omc/2\pi=710$ nHz and $\omc/\ome=5$. A random noise with an rms of 15 nHz (top), 30 nHz (middle), and 50 nHz (bottom) was added to the mode frequencies. The colors correspond to the different scenarios that were tested: ideal case where all the components are detected (black), inclination angle of 90$^\circ$ (purple) or 30$^\circ$ (red), and case of an overlap with the closest radial mode (blue). The vertical dashed lines indicate the actual values of $\omc$ and $\ome$ computed with Eq. \ref{eq_omc} and \ref{eq_ome}.
\label{fig_om_histo}}
\end{figure*}

We observe from Fig. \ref{fig_omc_ome} that the recovered values of $\omc$ and $\ome$ are highly correlated. This can be understood because the average rotation $\omtot$ over the whole interior, which can be written as
\begin{linenomath*}
\begin{equation}
\omtot \equiv \frac{\int_0^R K(r)\Omega(r)\,\hbox{d}r}{\int_0^R K(r)\,\hbox{d}r} = \frac{\beta_{\rm c}\omc+\beta_{\rm e}\ome}{\beta_{\rm c}+\beta_{\rm e}},
\label{eq_omtot}
\end{equation}
\end{linenomath*}
can be recovered from the observed modes frequencies much more precisely than $\omc$ and $\ome$ individually. As a result, the quantity ${\beta_{\rm c}\omc+\beta_{\rm e}\ome}$ is expected to be roughly conserved for all the iterations of the Monte Carlo simulation. The conservation of this quantity in the $(\ome,\omc)$ plane defines a straight line, which is overplotted in Fig. \ref{fig_omc_ome} and accounts well for the correlation between the fitted values of $\omc$ and $\ome$.

To study the influence of the uncertainties of mode frequencies on the performance of our method, we performed Monte Carlo simulations considering error bars of 30 and 50 nHz. The distributions of the recovered values of $\omc$ and $\ome$ are shown in Fig. \ref{fig_om_histo} (black histograms). The error bars of $\omc$ and $\ome$ increase to $\sim 35$ nHz for $\sigma_\nu=30$ nHz and to $\sim 65$ nHz for $\sigma_\nu=50$ nHz. In the latter case, the distributions of $\omc$ and $\ome$ become slightly bimodal, with a small secondary peak at $\omc/2\pi\sim480$ nHz and $\ome/2\pi\sim360$ nHz. These spurious solutions are obtained along with extremum values of $\Delta t$ (either 0 Myr or 8 Myr), i.e., away from the avoided crossing. These solutions therefore have negligible multiplet asymmetries, and they correspond to the solutions that would be obtained if near-degeneracy effects were ignored. The fact that the fits sometimes converge to this spurious solution in our noisiest scenario can be understood because the frequency uncertainty (50 nHz) becomes comparable to the asymmetry of the multiplets (the numerator of $\delta_{\rm asym}$ is around 135 nHz for the rotation profile that we have chosen for these simulations). If we use rotation profiles with larger core-envelope contrasts, the asymmetry of rotational multiplets during the avoided crossing increases and the bimodality of the recovered rotation rates vanishes, even for $\sigma_\nu=50$ nHz.

In our Monte Carlo simulations, the parameter $\Delta t$ was also recovered with very little scatter ($3.63\pm0.02$ for $\sigma_\nu=15$ nHz). As was done before, we used the optimal value of $\Delta t$ to estimate $\zeta_{0,a}$ and $\zeta_{0,b}$ for each simulation. The distribution obtained for $\zeta_{0,a}$ is shown in Fig. \ref{fig_zeta_histo} (the distribution of $\zeta_{0,b}$ is qualitatively similar). The recovered values are in close agreement with the actual values of the $\zeta$ parameters computed by inserting the eigenfunctions of model D into Eq. \ref{eq_zeta} (we obtained $\zeta_{0,a}=0.481\pm0.008$ and $\zeta_{0,a}=0.545\pm0.008$ for $\sigma_\nu=15$ nHz). This confirms that the trapping of the modes can be precisely recovered, even when the effects of noise are included.

\subsubsection{Effects of missing components \label{sect_missing_m}}

In addition to the effects of the noise, we must take into account the fact that in general not all the components of the $l=2$ multiplets can be detected. This is caused mainly by the inclination of the star, but also by the possible overlap of certain components with the adjacent radial or mixed dipolar modes. To determine the effects of missing components on the fitting procedure, we tested the following scenarios:
\begin{enumerate}
\item Ideal case (all $m$-components are detected)
\item Inclination angle of 90$^\circ$ ($m=\pm1$ components missing)
\item Inclination angle of $\sim 30^\circ$ ($m=\pm2$ components missing)
\item Overlap of a part of the higher-frequency multiplet with the closest radial mode ($m=+1,\,+2$ components of this multiplet missing)
\end{enumerate} 

\begin{figure}
\begin{center}
\includegraphics[width=9cm]{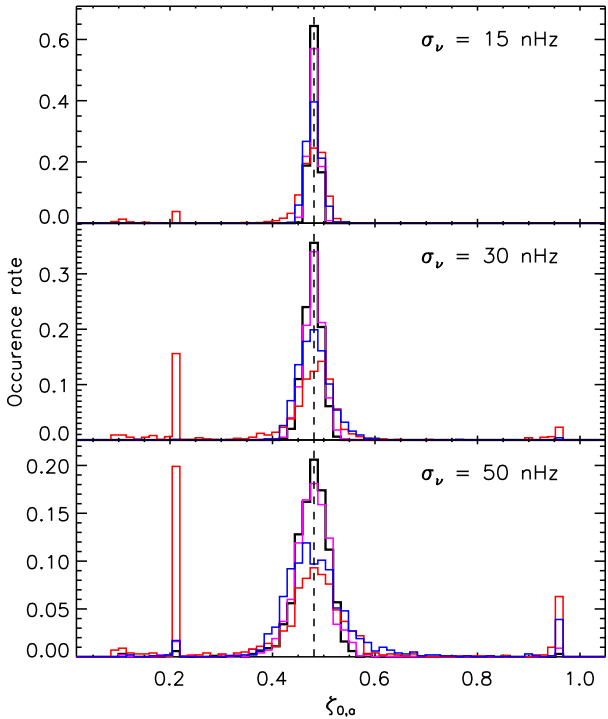}
\end{center}
\caption{Distribution of the values of $\zeta_{0,a}$ recovered when fitting Eq. \ref{eq_pert_omce} to the mode frequencies of \textsc{ACOR} for model D, assuming a rotation profile with $\omc/2\pi=710$ nHz and $\omc/\ome=5$. The lines and colors have the same meaning as in Fig. \ref{fig_om_histo}. The vertical dashed lines indicate the actual value of  $\zeta_{0,a}$ computed with the eigenfunctions of the model and Eq. \ref{eq_zeta}.
\label{fig_zeta_histo}}
\end{figure}

Scenario~1 was already tested in Sect. \ref{sect_noise}. We repeated the same Monte Carlo simulations for scenarios 2, 3, and 4. The distributions of the obtained values of $\omc$ and $\ome$ are shown in Fig. \ref{fig_om_histo}. For scenarios 2 and 4, the missing components result in a slight increase in the error bars of the recovered core and envelope rotation rates, but overall the performance of the method remains quite comparable to the ideal case where all the components are detected. In the case of a low inclination angle (scenario 3), the performance of the method is much poorer, with the bimodality of the distributions of $\omc$ and $\ome$ appearing even with $\sigma_\nu = 15$ nHz. This is expected since in the absence of the $m=\pm2$ components, the asymmetry can only be detected through the $m=\pm1$ components and it is much smaller. With the rotation profile that we have chosen here, $\omega_{m=-1}+\omega_{m=1}-2\omega_{m=0}$ is around 35 nHz, which is comparable to the uncertainties of the mode frequencies. This explains the large increase of the secondary peak in the distributions for scenario 3 as $\sigma_\nu$ increases. An mentioned in Sect. \ref{sect_noise}, this secondary peak corresponds to the cases where the fit converged toward $\Delta t=0$ Myr or $\Delta t=8$ Myr, i.e. toward a configuration without multiplet asymmetries. These cases are clearly identified in Fig. \ref{fig_zeta_histo} because they form sharp peaks at $\zeta_{0,a}\sim0.2$ and $\zeta_{0,a}\sim 0.96$, which correspond to $\Delta t=8$ Myr or $\Delta t=0$ Myr, respectively.

We therefore conclude that provided the inclination angle of the star is large enough to detect $m=\pm2$ components ($i\gtrsim50^\circ$), the full set of \kepler\ data (four years of observations) should be sufficient to reliably estimate the mean core and envelope rotation rates using only two $l=2$ modes undergoing an avoided crossing.

\section{Application to KIC7341231 \label{sect_otto}}

The young red giant KIC7341231 is a favorable case for study considering the results of our Monte Carlo simulation. The inclination angle of the star is high ($i=85^\circ\pm5^\circ$, \citealt{deheuvels12}) so that the $m=\pm2$ components of the $l=2$ multiplets are clearly visible. Compared to scenario 2 which was tested in Sect. \ref{sect_missing_m}, we also have measurements for the frequencies of the $m=+1$ components, even though the associated error bars are larger than for the other components. The error bars of the $m=-2,0,$ and 2 components are all around or below 15 nHz. Consequently, we expected to be able to reliably extract $\omc$ and $\ome$ for this star using our method.

We fitted Eq. \ref{eq_pert_omce} to the observed frequencies of the two $l=2$ multiplets which are listed in Table \ref{tab_otto_freq}. The fit converged to a core rotation of $\omc/2\pi=771\pm13$ nHz and an envelope rotation of $\ome/2\pi=45\pm12$ nHz. The values obtained for the other fitted parameters are given in Table \ref{tab_res_otto}. The quoted error bars were obtained by performing Monte Carlo simulations using the observed mode frequencies and their uncertainties. We repeated the fitting procedure using a wide range of initial parameters. We found that if $\omc>\ome$ in the set of initial parameters, we always recover the solution that is given in Table \ref{tab_res_otto}, regardless of the initial guess for $\Delta t$. As mentioned in Sect. \ref{sect_simu_noise}, we found that if the fit is started with initial rotation rates such that $\ome>\omc$, it converges to a solution with the envelope rotating much faster than the core ($\omc/2\pi=67\pm14$ nHz and $\ome/2\pi=675\pm12$ nHz, with $\Delta t=3.98\pm0.02$). This alternate solution can be clearly excluded for KIC7341231 because if it were the case, the g-like dipolar modes would have smaller splittings than the p-like ones, contrary to what is observed. 

The mode frequencies produced by the set of best-fit parameters are overplotted in Fig. \ref{fig_avcross_l2_otto} (vertical dotted lines). They are in very close agreement with the observed mode frequencies. The agreement can be qualitatively assessed by computing the $\chi^2$ using the observed frequencies and error bars listed in Table \ref{tab_otto_freq}. For the best-fit parameters, we obtained $\chi^2=4.8$. This translates into a reduced $\chi^2$ of 1.6 (the fit involves 5 free parameters for 8 observables, thus yielding a number of degrees of freedom of 3), which indicates a statistically good agreement. With this procedure, we were also able to predict the location of the $m=-1$ components of the multiplets, which could not be detected from the observed oscillation spectrum. For the lower-frequency multiplet, the theoretical location of the $m=-1$ component coincides with a slight excess of power, while nothing can be seen in the power spectrum around the frequency predicted for the $m=-1$ component of the higher-frequency multiplet. 

\begin{table}
  \begin{center}
  \caption{Optimal parameters resulting from the fit of Eq. \ref{eq_pert_omce} to the observed frequencies of the two $l=2$ multiplets in avoided crossing in the spectrum of KIC7341231. \label{tab_res_otto}}
\begin{tabular}{l c} 
\hline \hline
\T $\omc/2\pi$ (nHz) &  $771\pm13$ \\
$\ome/2\pi$ (nHz) & $45\pm12$ \\
$\omega_{0,a}/2\pi$ ($\mu$Hz) & $380.366\pm0.009$ \\
$\omega_{0,b}/2\pi$ ($\mu$Hz) & $382.619\pm0.009$ \\
$\Delta t$ (Myr) & $4.14\pm0.02$ \\
$\zeta_{0,a}$ & $0.304\pm0.006$ \\
\B $\zeta_{0,b}$ & $0.722\pm0.006$ \\
\hline
\end{tabular}
\end{center}
\end{table}

%

\section{Discussion \label{sect_discussion}}

\subsection{Other possible sources of multiplets asymmetry \label{sect_other}}

As we have shown in this paper, near-degeneracy effects are very likely the origin of the multiplet asymmetries in KIC7341231. Conditions that cause near-degeneracy of modes are expected to occur in this star, and our treatment of these effects shows that they produce asymmetries similar to those that are observed. Additionally, no asymmetries were found in the $l=1$ multiplets of KIC7341231, which implies that whatever phenomenon is producing multiplet asymmetries in this star does not affect dipolar modes. This can be accounted for by near-degeneracy effects, because the evanescent region for $l=1$ modes is thinner, which produces a much stronger coupling between the g-mode and p-mode cavities. As a result, the frequency separation between dipolar mixed modes is larger than for quadrupolar modes and near-degeneracy effects are much weaker. Based on \cite{deheuvels12}, the smallest frequency spacing between $l=1$ modes in the spectrum of KIC7341231 ($7.64\pm0.05\,\mu$Hz) represents more that 10 times the core rotation rate of the star, which confirms that near-degeneracy-effects should be small for dipolar modes in this star. For the sake of completeness, we briefly mention other potential sources of asymmetries.

Asymmetries in rotational multiplets can be the signature of a rotation rate that is fast enough to produce non-negligible effects of the centrifugal force and second-order effects of the Coriolis force. In this case, higher-order perturbations need to be included in the oscillation equations, which break the symmetry of rotational multiplets (e.g., \citealt{dziembowski92}, \citealt{suarez06}). However, the rotation rate estimated for KIC7341231 by \cite{deheuvels12} is much too low for second-order effects of rotation to significantly contribute to rotation splittings. Besides, no asymmetries were detected in the $l=1$ multiplets of KIC7341231, contrary to what would be expected in the presence of fast rotation.

It is also known that latitudinal differential rotation can produce asymmetric rotational multiplets. To first-order, the frequency shifts caused by rotation can be written as
\begin{linenomath*}
\begin{equation}
\delta\omega_{n,l,m} = m \int_0^R\int_0^\pi K_{n,l,m}(r,\theta)\Omega(r,\theta)\,\hbox{d}r\hbox{d}\theta
\end{equation} 
\end{linenomath*}
Since the kernels $K_{n,l,m}(r,\theta)$ depend on the azimuthal order $m$, the splittings of the different $m$ components, usually computed as $(\omega_{-m}-\omega_{m})/2m$, differ. However, it can be readily shown from the expression of the kernels that $K_{n,l,-m}=K_{n,l,m}$, so that $\delta\omega_{n,l,-m}=-\delta\omega_{n,l,m}$ (e.g. \citealt{aerts10}). As a result, the asymmetry as defined by $\delta_{\rm asym}$ in Eq. \ref{eq_asym} is expected to cancel \avoir{at first-order} in presence of latitudinal differential rotation. \avoir{Second-order effects of rotation may on the other hand produce non-zero values of $\delta_{\rm asym}$. In order to rule out latitudinal differential rotation as the cause of the observed asymmetries, we performed non-perturbative calculations with \textsc{ACOR}. The rotation law was chosen differential in latitude only, of the form $\Omega(\theta) = \Omega_{\rm eq} ( 1 - \Delta \Omega / \Omega_{\rm eq} \cos \theta^2)$, where $ \Omega_{\rm eq}$ and $ \Omega_{\rm pol}$ are the equatorial and polar rotation, respectively, and $\Delta \Omega \equiv \Omega_{\rm eq} - \Omega_{\rm pol}$. We chose $\Omega_{\rm eq}/2\pi$ = 710 nHz, and a latitudinal differential rotation of $\Delta \Omega / \Omega_{\rm eq}=0.36$, which corresponds to the solar value. With this rotation profile, we obtained asymmetries of the order of $\delta_{\rm asym}\sim10^{-3}$, i.e. much lower than the observed values}. We thus conclude that latitudinal differential rotation cannot account on its own for the observed asymmetries.

Finally, multiplet asymmetries can also be the signature of an internal magnetic field. The possibility of a buried magnetic field inclined with respect to the rotation axis can be excluded for KIC7341231, since such a field would produce a magnetic splitting of each of the $2l+1$ components in each rotational multiplets (\citealt{dicke82}, \citealt{dziembowski85}), and this is not observed. However, an internal field aligned with the rotation axis would not change the number of components, but alter the fine structure pattern of the multiplets, potentially producing asymmetries. Establishing the impact of such a field on the rotational multiplets would require a dedicated study and is out of the scope of the present paper. We can however remark that a buried magnetic field would produce more distortion in the g-dominated multiplets than in the p-dominated ones (\citealt{hasan92}), which is the opposite of what is observed in KIC7341231 and KIC5006817. Also, it is hard to imagine how an internal magnetic field would produce asymmetries in quadrupolar multiplets but not in dipolar ones in KIC7341231. 

\subsection{Near-degeneracy effects in red giants}

In this paper, we have focused on the case of the young red giant KIC7341231. The internal rotation of this star is not peculiar compared to other similar targets, thus near-degeneracy effects are expected to be important for many other giants. In fact, since the intensity of the coupling between the p-mode and g-mode cavities decreases during the evolution (e.g. \citealt{mosser17}), we anticipate that near-degeneracy effects become even larger for giants more evolved than KIC7341231, which lies at the base of the RGB. Therefore, these effects should always be considered when using the rotational multiplets of quadrupolar modes to probe the internal rotation of red giants.

For dipolar modes, near-degeneracy effects are smaller than those for quadrupolar modes, as explained in Sect. \ref{sect_other}. However, they need to be considered for giants slightly more evolved than KIC7341231. This is the case for KIC5006817, for which asymmetrical $l=1$ multiplets have been reported (\citealt{beck14}). It is straightforward to show that near-degeneracy effects are expected in this star. The closest frequency separation between $l=1$ mixed modes in this star is $\sim 0.8\,\mu$Hz, which is even smaller than the core rotation rate of 0.93 $\mu$Hz inferred for this star by \cite{beck14}. In fact, the asymmetries detected in this star show a lot of similarities with the ones produced by near-degeneracy effects: they arise only for the modes that show the highest level of mixing (see Fig. 6 of \citealt{beck14}) and it is clear from Fig. 5 of \cite{beck14} that for all pairs of asymmetric multiplets, the lowest-frequency multiplet has a negative asymmetry, while the largest-frequency multiplet has a positive asymmetry, as we have found in Sect. \ref{sect_pert}. \avoir{In the case of KIC4448777, for which \cite{dimauro16} reported asymmetries in the multiplets of several $l=1$ modes, the authors found a mean core rotation of 0.75 $\mu$Hz, which represents about 40\% of the closest frequency spacing between $l=1$ mixed modes. We thus expect near-degeneracy effects to be non-negligible in this star also, although a dedicated study would be necessary to determine whether they are the cause of the reported asymmetries.}

\section{Conclusion \label{sect_conclusion}}

In this paper, we investigated the origin of the asymmetries that have been detected in the rotational multiplets of mixed modes in red giants. We showed that in both red giants in which asymmetrical multiplets have been reported (KIC7341231, \citealt{deheuvels12}; KIC5006817, \citealt{beck14}), near-degeneracy effects are expected to arise because of the small frequency spacing between mixed modes. Such effects had not been included in previous studies. Using the young red giant KIC7341231 as a test case, we accounted for the effects of near-degeneracy in the framework of a first-order perturbative approach and showed that these effects indeed produce multiplet asymmetries with very similar features compared to the observations. We also validated this perturbative approach using the code \textsc{ACOR} (\citealt{ouazzani12}), which solves the oscillation equations including the effects of rotation in a non-perturbative manner.

We showed that the asymmetries of rotational multiplets arising for near-degenerate mixed modes are linked to differential rotation between the core (g-mode cavity) and the envelope (p-mode cavity). Indeed, for near-degenerate mixed modes, the components of the rotational multiplets are trapped differently inside the star. Each component is thus sensitive to an average of the internal rotation with different weights given to the core and the envelope. By exploiting this property, we developed a method to measure the mean core and envelope rotation using the rotational multiplets of near-degenerate mixed modes. This method was then tested using Monte Carlo simulations on artificial data, where we included the effects of the uncertainties of mode frequencies and considered different scenarios for the mode visibilities within the multiplets. We showed that provided the dataset is long enough for the asymmetries to be detected at a statistically significant level, our method can be used to successfully and precisely recover the mean rotation rates of the core and envelope using the frequencies of only two $l=2$ near-degenerate rotational multiplets. Higher-inclination angles are more favorable, because then the $m=\pm2$ components can be detected, and they are the ones that show the largest asymmetries. 

Next, we successfully applied our method to the young red giant KIC7341231. We obtained a core rotation rate of $\omc/2\pi=771\pm13$ nHz for this star. This measurement is in good agreement with the results of \cite{deheuvels12}, who had obtained $\omc/2\pi=710\pm51$ nHz by applying the OLA inversion method to the dipolar modes. The improved precision on the measurement of $\omc$ is likely related to the longer time series that was used here compared to \cite{deheuvels12}, who had only one year of data at their disposal. This result shows that we are now able to measure the core rotation rate using quadrupolar mixed modes, which had not been possible earlier. We intend to investigate in a future work whether this measurement, combined to the one obtained with dipolar modes, can be used to place constraints on the  latitudinal differential rotation of the core --- we know that modes of different degree are sensitive to different latitudes of the stellar rotation profile.

Interestingly, our method also enabled us to provide an estimate of the envelope rotation of KIC7341231, which had not been possible for this star using the dipolar modes. We obtained $\ome/2\pi=45\pm12$ nHz, which is in agreement with the upper limit given by \cite{deheuvels12}. We thus obtained an estimate of the core-envelope contrast of $17\pm5$ for KIC7341231. This is an improvement compared to the results of \cite{deheuvels12} who could only provide a lower limit of 5 for the core-envelope rotation ratio. This measurement could be used to improve our knowledge of the timescale over which angular momentum is transported in red giants (\citealt{eggenberger17}).

Based on our results, we expect near-degeneracy effects to be important for quadrupolar mixed modes all along the red-giant branch. For dipolar mixed modes, these effects are negligible at the base of the red-giant branch, but must be taken into account for more evolved giants. We have shown in this paper that rotational multiplets affected by near-degeneracy effects require special treatment and that they give us information about the internal rotation of red giants.

\begin{acknowledgements}
We would like to thank T. Appourchaux who initially suggested us to look into the role of rotation on avoided crossings. We are also thankful to M.~A. Dupret, J. Ballot, M~J. Goupil, and B. Mosser for fruitful discussions that were very helpful. We also thank the anonymous referee for suggestions that improved the clarity of the paper. SD also wishes to thank the KITP at UCSB for hosting the research program ``Asteroseismology in the Space Age'', in this work started. SD acknowledges support from the PNPS under the grant ``Rotation interne et magn\'etisme des sous-g\'eantes et g\'eantes Kepler'' and from the Centre National d'\'Etudes Spatiales (CNES). Funding for the Stellar Astrophysics Centre is provided by The Danish National Research Foundation (Grant agreement no.: DNRF106). The research is supported by the ASTERISK project (ASTERoseismic Investigations with SONG and Kepler) funded by the European Research Council (Grant agreement no.: 267864). RMO acknowledges financial support by a CNES post-doctoral fellowship. SB acknowledges NASA grant NNX16AI09G and NSF grant AST-1514676.
\end{acknowledgements}

\bibliographystyle{aa.bst} 
\bibliography{biblio} 

\begin{appendix}

\section{First-order perturbative approach including near-degeneracy effects \label{app_neardeg}}

Let us consider the case of two modes (indicated by subscripts $a$ and $b$) with near-degenerate eigenfrequencies, i.e., $|\omega_{0, a}-\omega_{0, b}|\sim\Omega$. In this case, the eigenfunctions of the near-degenerate modes can be written as $\vect{\xi} = A\vect{\xi}_{0, a} + B\vect{\xi}_{0, b}$. By injecting this expression in Eq. \ref{eq_pert} and taking the inner product with $\vect{\xi}_{0, a}$ and $\vect{\xi}_{0, b}$ successively, the following system is obtained
\begin{linenomath*}
\begin{align}
\left(\omega_{0,a}^2 - \omega^2 + 2\omega_{0,a}\omega_{1,a}\right) A + 2\omega_{0,a}\omega_{1,ab}B & = 0 \nonumber \\
2\omega_{0,b}\omega_{1,ab}A  +  \left(\omega_{0,b}^2 - \omega^2 + 2\omega_{0,b}\omega_{1,b} \right)B & = 0 \label{eq_syst_AC}
\end{align}
\end{linenomath*}
where we have defined 
\begin{linenomath*}
\begin{equation}
\omega_{1,ab} \equiv \frac{1}{2\omega_0} \langle \mathcal{L}_1 \vect{\xi}_{0,a} | \vect{\xi}_{0,b} \rangle
\end{equation}
\end{linenomath*}
To first order, one can write $\omega_{0,a}^2+ 2\omega_{0,a}\omega_{1,a}-\omega^2\sim2\omega_{0,a}(\omega_a-\omega)$, where $\omega_a$ corresponds to the first-order perturbed frequency of mode $a$ in the absence of near-degeneracy (see Eq. \ref{eq_omab}) and a similar expression holds for mode $b$. Eq. \ref{eq_syst_AC} can thus be rewritten as
\begin{linenomath*}
\begin{equation}
\begin{pmatrix}
\omega_{a}-\omega & \omega_{1,ab} \\
\omega_{1,ab} & \omega_{b}-\omega \\
\end{pmatrix}
\begin{pmatrix}
A \\
B \\
\end{pmatrix}
=
\begin{pmatrix}
0 \\
0 \\
\end{pmatrix}
\label{eq_syst_AC2}
\end{equation}
\end{linenomath*}
Solutions to Eq. \ref{eq_syst_AC2} can exist only if 
\begin{linenomath*}
\begin{equation}
(\omega_a-\omega)(\omega_b-\omega) - \omega_{1,ab}^2 = 0
\end{equation}
\end{linenomath*}
This yields the following expression for the frequencies of the near-degenerate modes
\begin{linenomath*}
\begin{equation}
\omega_\pm = \frac{\omega_{a}+\omega_{b}}{2} \pm \frac{1}{2} \sqrt{(\omega_{a}-\omega_{b})^2 + 4\omega_{1,ab}^2}
\end{equation}
\end{linenomath*}
Eq. \ref{eq_syst_AC2} can also be used to obtain a relation between $A$ and $B$, and therefore to estimate the eigenfunctions of the perturbed modes.

\section{An approximate expression for $\omega_{1,ab}$ \label{app_omega1ab}}

We here obtain an approximate expression for the coupling term $\omega_{1,ab}$, starting from Eq. \ref{eq_omega1ab}, in order to interpret its behavior as a function of the rotation profile. For this purpose, we introduce the quantity 
\begin{linenomath*}
\begin{equation}
\gamma\equiv\int_0^R K_{ab}(r)\,\hbox{d}r,
\label{eq_gamma}
\end{equation}
\end{linenomath*}
where $K_{ab}$ is defined by Eq. \ref{eq_Kab}. Taking into account the fact that the unperturbed eigenfunctions of modes $a$ and $b$ are orthogonal, Eq. \ref{eq_gamma} can be rewritten
\begin{linenomath*}
\begin{equation}
\gamma = - \int_0^R \rho_0 r^2 \left( \xi_{{\rm h},0,a}\xi_{{\rm h},0,b} + \xi_{{\rm r},0,a}\xi_{{\rm h},0,b} +\xi_{{\rm h},0,a}\xi_{{\rm r},0,b} \right)  \,\hbox{d}r
\end{equation}
\end{linenomath*}
In the g-mode cavity, the horizontal displacement is much larger than the vertical displacement for all modes, so that the cross-terms $\xi_{\rm r}\xi_{\rm h}$ can be safely neglected in this region. In the p-mode cavity, these cross-terms dominate, but since the horizontal displacement is much smaller that the radial displacement, the overall contribution of the p-mode cavity to $\gamma$ is smaller than the contribution of the g-mode cavity. We can thus approximately write
\begin{linenomath*}
\begin{equation}
\gamma \approx - \int_0^{r_{\rm c}} \rho_0 r^2 \xi_{{\rm h},0,a}\xi_{{\rm h},0,b} \,\hbox{d}r
\end{equation}
\end{linenomath*}
In the g-mode cavity, considering that $\xi_{\rm h}\gg\xi_{\rm r}$, the quantity $\gamma_{\rm c}$ (defined by Eq. \ref{eq_gammac}) can be approximated by
\begin{linenomath*}
\begin{align}
\gamma_{\rm c} & \approx \int_0^{r_{\rm c}} \rho_0 r^2 (L^2-1) \xi_{{\rm h},0,a}\xi_{{\rm h},0,b}  \,\hbox{d}r \\
& \approx (1-L^2)\gamma 
\end{align}
\end{linenomath*}
Hence
\begin{linenomath*}
\begin{equation}
\gamma_{\rm e} = \gamma - \gamma_{\rm c} \approx L^2 \gamma
\end{equation}
\end{linenomath*}
Consequently, Eq. \ref{eq_omega1ab} can be rewritten as
\begin{linenomath*}
\begin{equation}
\omega_{1,ab} \approx m \gamma \left[ (1-L^2)\omc + L^2 \ome \right] 
\label{eq_omega1ab_approx}
\end{equation}
\end{linenomath*}
In the particular case of a rotation profile for which the rotational splitting of pure g modes is identical to that of pure p modes, one has $\omc = L^2/(L^2-1) \ome$, and it can be seen from Eq. \ref{eq_omega1ab_approx} that $\omega_{1,ab}$ vanishes.

\end{appendix}

\end{document}